\documentclass[sigconf]{aamas} 

\usepackage{xcolor}
\usepackage{subfigure} 
\usepackage{amsmath}
\usepackage{ulem}

\usepackage[T1]{fontenc}

\usepackage{aecompl}
\usepackage{fancyhdr}
\usepackage{newfloat}
\usepackage{listings}
\usepackage{balance} 
\makeatletter

\newcommand{\Rmnum}[1]{\expandafter\@slowromancap\romannumeral #1@}
\makeatother

\usepackage{algorithm}  
\usepackage{algorithmicx}  
\usepackage{algpseudocode}
\usepackage{makecell}

\usepackage{bbm}
\usepackage{dsfont}

\setcopyright{ifaamas}
\acmConference[AAMAS '23]{Proc.\@ of the 22nd International Conference
on Autonomous Agents and Multiagent Systems (AAMAS 2023)}{May 29 -- June 2, 2023}
{London, United Kingdom}{A.~Ricci, W.~Yeoh, N.~Agmon, B.~An (eds.)}
\copyrightyear{2023}
\acmYear{2023}
\acmDOI{}
\acmPrice{}
\acmISBN{}
\pdfoutput=1  

\acmSubmissionID{371}

\title[AC2C]{AC2C: Adaptively Controlled Two-Hop Communication for Multi-Agent Reinforcement Learning}

\author{Xuefeng Wang $^*$}
\affiliation{
  \institution{The Hong Kong University of Science and Technology}
  \city{Hong Kong}
  \country{China}}
\email{eexuefengw@ust.hk}

\author{Xinran Li $^*$}
\affiliation{
  \institution{The Hong Kong University of Science and Technology}
  \city{Hong Kong}
  \country{China}}
\email{xinran.li@connect.ust.hk}

\author{Jiawei Shao}
\affiliation{
  \institution{The Hong Kong University of Science and Technology}
  \city{Hong Kong}
  \country{China}}
\email{jiawei.shao@connect.ust.hk}

\author{Jun Zhang}
\affiliation{
  \institution{The Hong Kong University of Science and Technology}
  \city{Hong Kong}
  \country{China}}
\email{eejzhang@ust.hk}

\begin{abstract}
Learning communication strategies in cooperative multi-agent reinforcement learning (MARL) has recently attracted intensive attention. Early studies typically assumed a fully-connected communication topology among agents, which induces high communication costs and may not be feasible. 
Some recent works have developed adaptive communication strategies to reduce communication overhead, but these methods cannot effectively obtain valuable information from agents that are beyond the communication range.
In this paper, we consider a realistic communication model where each agent has a limited communication range, and the communication topology dynamically changes. 
To facilitate effective agent communication, we propose a novel communication protocol called \textit{Adaptively Controlled Two-Hop Communication} (AC2C). After an initial local communication round, AC2C employs an adaptive two-hop communication strategy to enable long-range information exchange among agents to boost performance, which is implemented by a communication controller. This controller determines whether each agent should ask for two-hop messages and thus helps to reduce the communication overhead during distributed execution.
We evaluate AC2C on three cooperative multi-agent tasks, and the experimental results show that it outperforms relevant baselines with lower communication costs.
\end{abstract}


\newcommand{\BibTeX}{\rm B\kern-.05em{\sc i\kern-.025em b}\kern-.08em\TeX}

\begin{document}


\pagestyle{fancy}
\fancyhead{}


\maketitle 


\section{Introduction}
\def\thefootnote{*}\footnotetext{These authors contributed equally to this work.}

Cooperative multi-agent reinforcement learning (MARL) \citep{busoniu2008comprehensive} has recently led to promising results in many real-world applications, such as robot control \citep{gupta2017cooperative} and autonomous driving \citep{shalev2016safe}. For example, in the path-finding task, MARL achieves similar performance as classic operation research algorithms but with much lower computational complexity \citep{sartoretti2019primal}. In the domain of games, well-trained agents have reached the master-level performance \citep{smac} and even won the game against professional players. 
In these applications, centralized training decentralized execution (CTDE) is a widely adopted paradigm due to its scalability potential and ability to deal with non-stationarity.

Classic CTDE architectures usually employ a centralized value network that leverages global information to guide agents' local policy training \citep{MADDPG}. During execution, each agent utilizes its local information to make decisions without centralized coordination. In practice, however, the partial observation and stochastic nature of MARL environments make it difficult for agents to accurately predict others' actions in such communication-free CTDE schemes, and thus miscoordination often happens. 

\begin{figure}[t] 
\centering 
\includegraphics[scale=0.33]{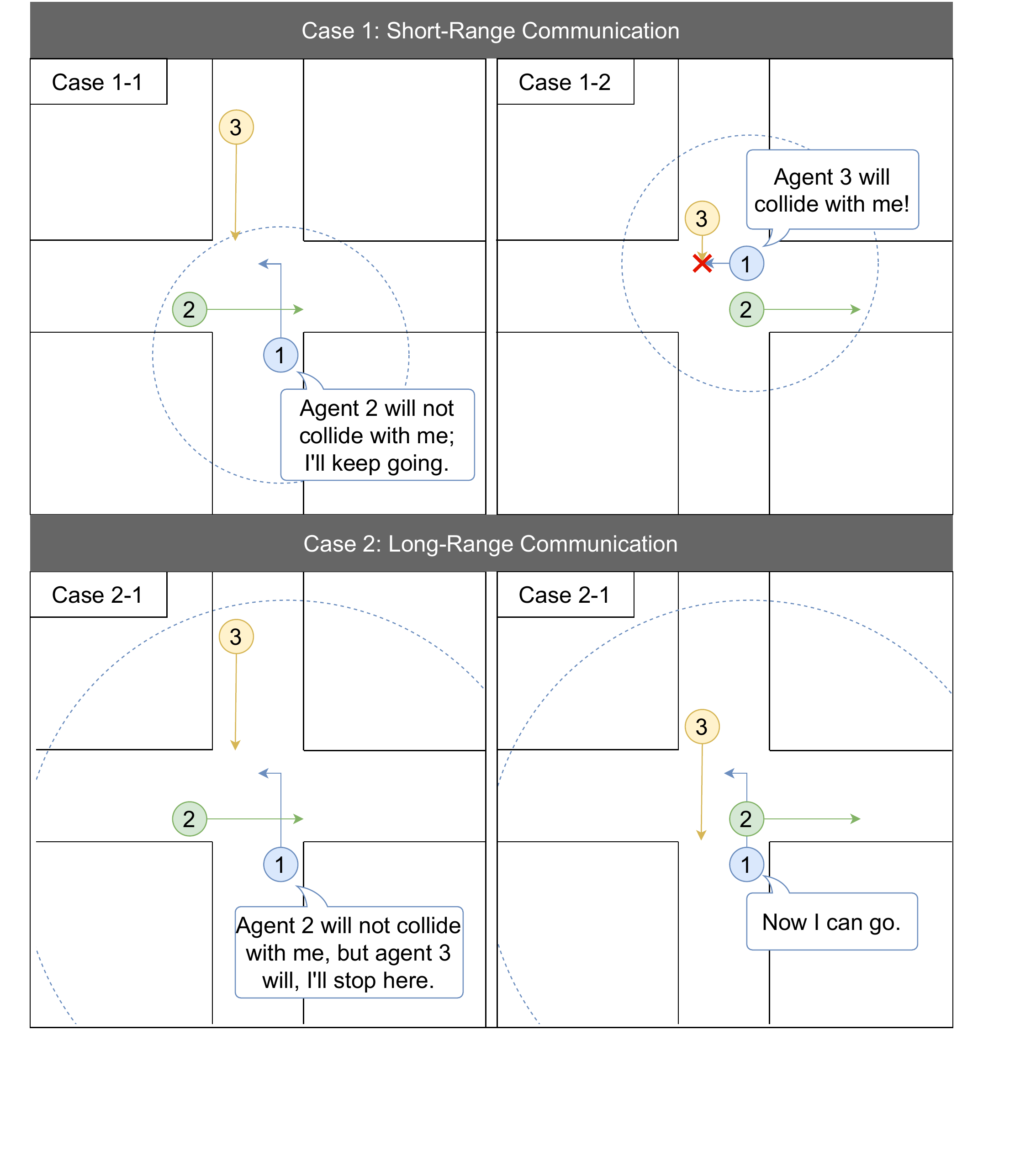} 
\caption{A didactic example in traffic junction environment. Consider agent $1$'s decision-making. In case 1, with only short-range communication, agent 1 only receives messages from agent 2, then it will enter the intersection and may collide with agent 3. In case 2, with both short-range and long-range communication, agent 1 is able to know agent $3$'s information in advance, therefore, it can make a better decision, i.e., waiting before entering the intersection.} 
\label{Fig_example} 
\end{figure}

To address this limitation, many recent studies on MARL enable communication to exchange information among agents in CTDE \citep{MADDPG, corr_commu}. These methods achieve superior performance compared with communication-free CTDE systems. Nevertheless, they often assume simplified communication models with a fully connected topology \citep{gcs}, i.e., each agent is able to receive a message from any other agent via point-to-point communication. 
In real systems, an agent's communication range is limited, e.g., when communicating over a wireless channel, and thus the communication topology among agents should be partially connected. Moreover, this topology will dynamically change as agents move. 
On the other hand, long-range information is desirable in MARL systems. As an example, consider a traffic junction task illustrated in Figure \ref{Fig_example}, where three cars need to pass through a traffic junction following predefined routes. In this case, obtaining long-range information helps agents plan ahead of time and avoid myopic decisions. Considering the realistic communication range constraint and the value of obtaining long-range information, an effective multi-hop communication mechanism is needed, which motivates our work. 

In this paper, we consider a cooperative MARL system, where agents are assumed to have limited communication ranges. To facilitate effective communication, we propose a novel communication protocol called \textit{Adaptively Controlled Two-Hop Communication} (AC2C). Enabled by an attention-based communication module and a multi-layer perceptron (MLP)-based controller, agents learn to adaptively engage in two-hop communication to balance the cooperative task performance and the communication overhead. 

The main contributions of this paper are summarized as follows:

\begin{itemize}
\item We consider a realistic MARL system with communication range constraints and propose a novel two-hop communication protocol, i.e., AC2C, to enable long-range information exchange.
\item Inspired by the gating mechanism \citep{mao2019learning}, we introduce an adaptive controller. This local controller adaptively determines whether the ego agent should ask for a subsequent communication round to obtain two-hop messages. In this way, expensive two-hop communication is only established whenever necessary.
\item We conduct experiments on three benchmark tasks, namely traffic junction, cooperative navigation, and predator prey. The superior performance compared with baselines demonstrates the effectiveness of our proposed method. 
We further analyze the communication costs in our experiments and show that AC2C achieves a good trade-off between communication cost and cooperative task performance.

\end{itemize}

\section{Related Work}
Recent studies \citep{MADDPG, QMIX, COMA} have made remarkable progress in MARL under the CTDE paradigm. Compared with its counterparts, i.e., independent learning \citep{IQL} and centralized training centralized execution (CTCE) \citep{CTCE}, CTDE has demonstrated significant performance and scalability potential. Nevertheless, communication-free CTDE exacerbates the partially observable issue and hinders effective cooperation, leading to sub-optimal decisions. 
Earlier works such as CommNet \citep{CommNet}, DIAL \citep{DIAL}, and BiCNet \citep{BiCNet} attempt to support communication in CTDE with a predefined topology. However, the performance of those methods often falls short in complex settings since relationships among agents constantly change in a dynamic multi-agent system, and a rigid communication graph cannot respond to such dynamics.

Subsequent works exploit state-dependent communication graphs to address the above shortcomings. In particular, ATOC \citep{atoc}, IC3Net \citep{ic3} and I2C \citep{i2c} introduce individual gating mechanisms to control the communication links among agents. The gating mechanisms are implemented with a classifier that determines whether to transmit messages based on local histories. Besides, VBC \citep{VBC} proposes a communication control unit depending on the local action confidence. The dynamically pruned communication graphs produced by these methods result in low communication costs. Nevertheless, the aforementioned works only consider single-round communication. In practical settings with limited communication ranges, agents cannot access information outside this range through single-round communication, which can severely limit the performance.

There have been some recent studies adopting multi-round communication to obtain more information from other agents. For instance, TarMAC \citep{Tarmac} utilizes multi-layer attention blocks to implement multi-round communication. But it requires a fully-connected communication topology, therefore causing prohibitive communication overhead. Furthermore, graph neural networks (GNNs) have recently been incorporated with MARL, owing to their power to enforce structural communication among agents. In particular,
DICG \citep{dicg} and DGN \citep{DGN} utilize graph convolutional networks (GCNs) to enable message passing among agents, while MAGIC \citep{magic} uses graph attention networks (GATs) to aggregate messages.
None of the above works dynamically prune the communication links (i.e. the edges in communication graphs), therefore they can be infeasible in realistic systems since densely connected communication graphs induce heavy communication overhead.
In this paper, we explicitly consider the effects of limited communication ranges on MARL systems. Our proposed method leverages multi-hop communication to enlarge the agents' reception fields. 
To reduce the communication cost, a decentralized controller is designed to determine whether an agent should request a subsequent communication round based on the obtained single-hop messages. In this manner, agents acquire the ability to obtain information outside their communication ranges and dynamically prune two-hop communication, thus achieving a good performance-communication trade-off.

\section{System Model}
\subsection{Problem Formation}
We formalize the problem as a decentralized partially observable Markov decision process (Dec-POMDP) \citep{pomdp}. It is modeled by a tuple $\mathcal{M} = \langle \mathcal{S}, A, P, R, \Omega, O, N, \gamma \rangle$, where $N$ is the number of agents, and $\gamma \in [0, 1)$ is the discount factor. At each timestep, the
environment state is $s \in \mathcal{S}$. Each agent $i$ receives a local observation $ o_i \in \Omega$ drawn from the observation function $O(s, i)$. Then, it selects an action $a_i \in A$, forming a joint action $\boldsymbol{a} \in A^N$, which leads to a next state $s'$ according to the transition function $P(s'| s, a)$. The agents collaboratively gain a global reward according to the reward function $r = R(s, \boldsymbol{a})$. Each agent keeps a local action-observation history at the current timestep $\tau_i  \in (\Omega \times A)$. The primary notations and descriptions are listed in Table \ref{table:notations}.

\subsection{Communication Protocol}
We consider a multi-agent system where agents are with limited communication ranges. For ease of illustration, we mainly consider distance-based communication constraints, i.e., an agent can only establish direct communication links with the ones within a range $L$. 

Formally, we call the neighboring agents that are located within distance $L$ to agent $i$ its one-hop neighbors, denoted by $\mathcal{N}_i^{(1)}$. And agents that are within the distance $L$ to any agent in the set $\mathcal{N}_i^{(1)}$ are denoted as $\mathcal{\widetilde{N}}_i^{(2)}$. We then define the set of agents belonging to $\mathcal{\widetilde{N}}_i^{(2)}$, excluding agent $i$'s one-hop neighbors and itself, as agent $i$'s two-hop neighbors, denoted as $\mathcal{N}_i^{(2)}= (\mathcal{\widetilde{N}}_i^{(2)} \backslash  \mathcal{N}_i^{(1)}) \backslash \{i\}$. 
For the example in Figure \ref{Fig_comm}, agents $2$ and $3$ are one-hop neighbors of agent $1$, while agents $4$ and $5$ are its two-hop neighbors. In the following, we first describe the GNN-based communication protocol adopted by existing studies and then introduce the proposed AC2C communication protocol.
\subsubsection{GNN-based Communication Protocol} 
GNN-based methods have recently been widely used for multi-agent communication \citep{dicg, DGN}. Before communication, agent $i$ holds a local feature embedding $c_i^{(0)}$. In the first communication round, it receives messages from its one-hop neighbors in $\mathcal{N}_i^{(1)}$ and updates it local embedding as $c_i^{(1)}$. In the second communication round, it communicates with nodes in $\mathcal{N}_i^{(1)}$ again, receiving their updated embeddings after the first communication round. In this way, each agent can obtain partial information from its two-hop neighbors, but in an indirect and inefficient sense. For example, in Figure \ref{Fig_comm}, agent 1 can obtain information from agents $4$ and $5$ after two communication rounds, but information on agents $6$ and $7$ is still unavailable.

\subsubsection{AC2C Communication Protocol}
\begin{figure}[t] 
\centering 
\includegraphics[width=6cm]{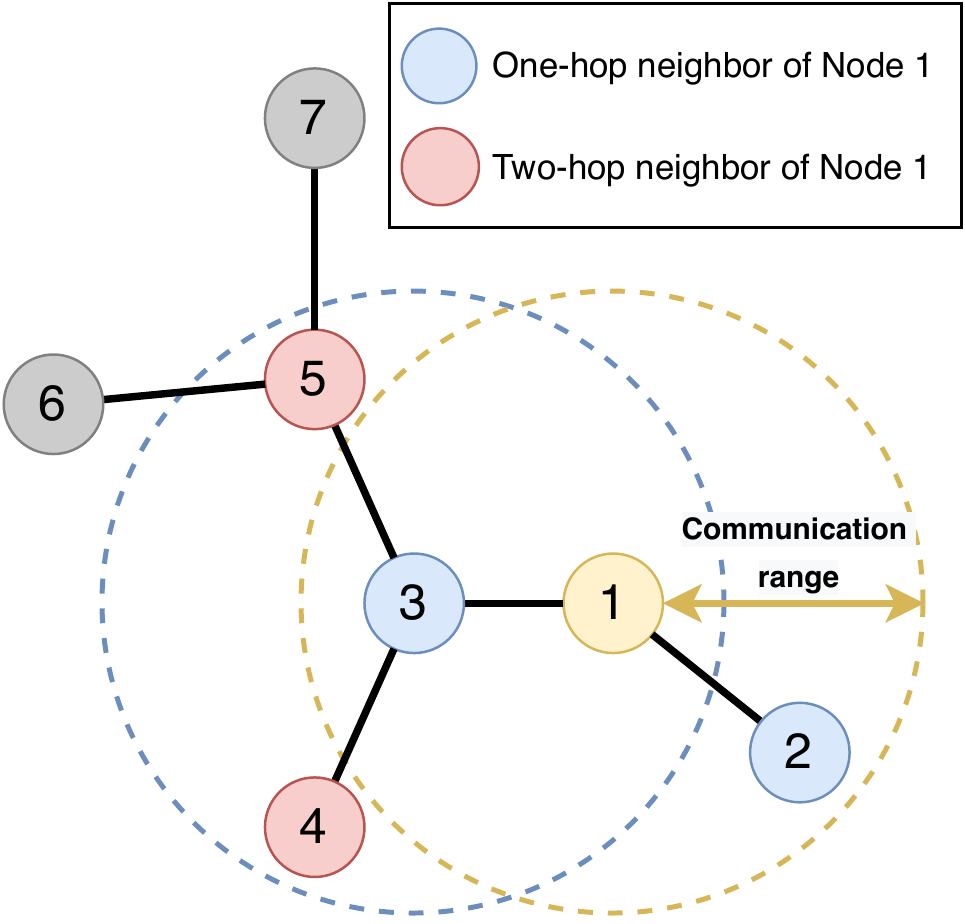} 
\caption{The two-hop communication model, where the blue and red nodes correspond to the one-hop and two-hop neighbors of node 1, respectively.
} 
\label{Fig_comm} 
\end{figure}

The proposed AC2C communication protocol adopts a two-hop communication mechanism to obtain information from two-hop neighbors $\mathcal{N}_i^{(2)}$, which facilitates effective communication among agents. During the first communication round, the communication process is the same as the GNN-based protocol, where each agent exchanges its local feature embedding $c^{(0)}_i$ with its one-hop neighbors. 
In the second communication round, however, agents exchange messages with their two-hop neighbors, while their one-hop neighbors only act as relay nodes. As illustrated in Figure \ref{Fig_comm}, the AC2C protocol can help agent 1 exploit information of agents $6$ and $7$ since agent $5$'s embedding (which contains partial information of agents 6 and 7) is transmitted to agent $1$ directly in the second communication round. Thus, AC2C can effectively enlarge agents' receptive fields compared to GNN-based communication protocol, leading to better performance. This performance gain comes with higher communication overhead, as it involves two-hop communication. Therefore, it is critical to adaptively control and reduce the frequency of evoking the expensive two-hop communication.

\section{Proposed Method}
\begin{figure}[t]
\begin{center}
\includegraphics[scale=0.48]{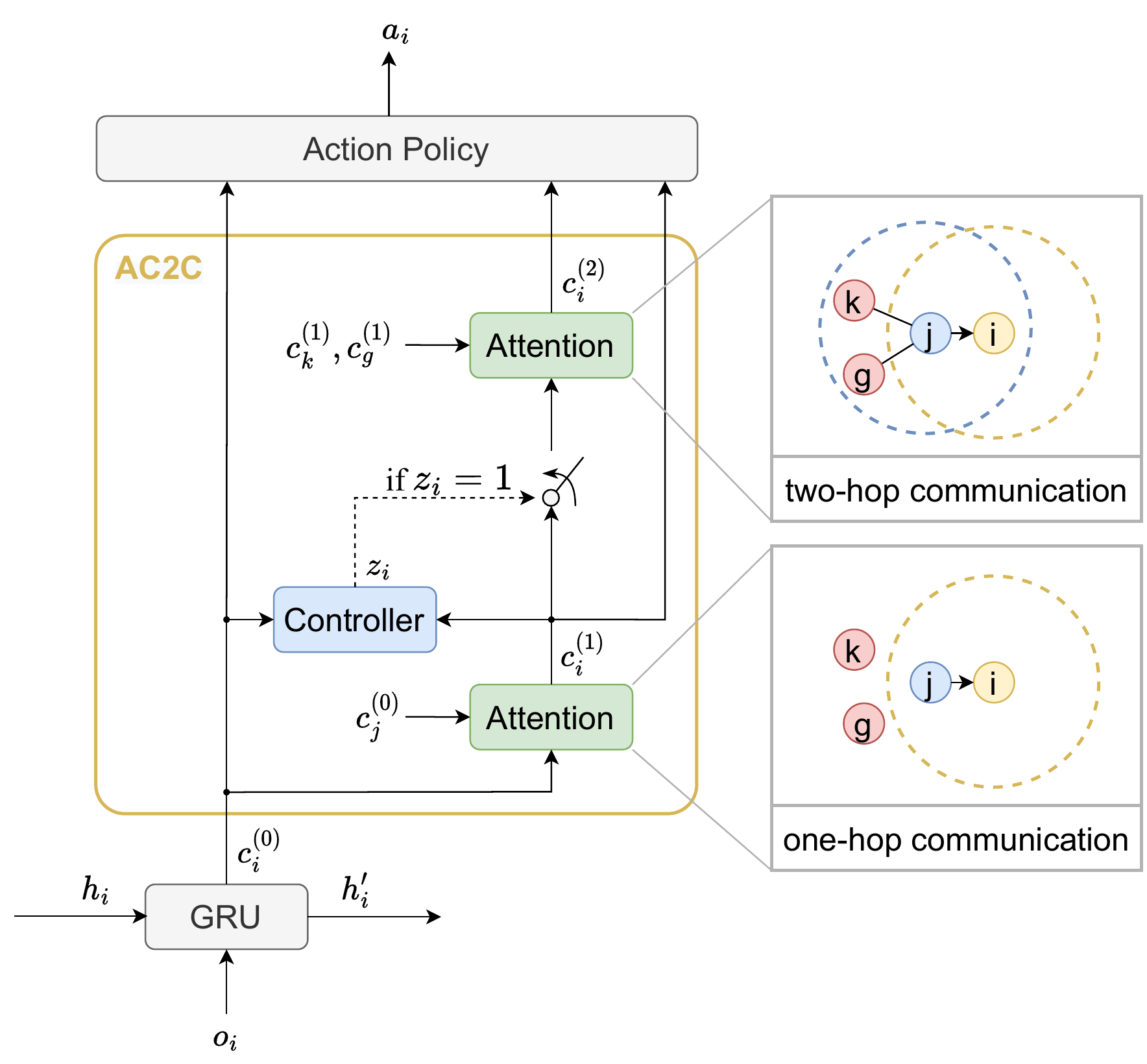}
\end{center}
\caption{
The architecture of the proposed AC2C network. At each timestep, agent $i$ receives a local observation $o_i$ and utilizes its historical information $h_i$ to generate an initial embedding $c_i^{(0)}$ and update its local historical information as ${h_i’}$. In the first communication round, agent $i$ receives a message from its one-hop neighbor $j$ and outputs the updated embedding $c_i^{(1)}$. 
Additionally, the local controller takes $c_i^{(0)}$ and $c_i^{(1)}$  as inputs and generates a binary signal $z_i$. If $z_i=0$, agent $i$ will not request the second communication round; if $z_i=1$, agent $i$ will request the second communication round. Upon receiving information from its two-hop neighbors $k$ and $g$ in the second communication round, agent $i$ aggregates messages again and produces $c_i^{(2)}$. After two communication rounds, agent $i$ generates an action based on $c_i^{(0)}$, $c_i^{(1)}$ and $c_i^{(2)}$ (if applicable).}
\label{AC2C_franework}
\end{figure}

In this work, we follow the conventional CTDE paradigm and augment it with the proposed AC2C communication protocol. In the proposed framework, each agent's local network consists of a GRU-based feature encoder, an AC2C communication module, and an MLP-based action policy network, as is shown in Figure \ref{AC2C_franework}. 

At each timestep, the feature encoder first takes the local observation $o_i$ as input to update its historical representation $h_i$ and outputs an initial local embedding $c_i^{(0)}$. Then, the agent leverages the local embedding and conducts a two-round communication process with the AC2C modules to exchange information with other agents, obtaining the updated local embedding $c_i^{(1)}$ and $c_i^{(2)}$ after each communication round. After that, the agents feeds $c_i^{(0)}$, $c_i^{(1)}$ and $c_i^{(2)}$ to the action policy network to generate the local action $a_i$.

\begin{table}[t]
\centering
\caption{Primary notations and descriptions.}
{
\begin{tabular}{c|c}
\hline
Notations & Description \\ 
\hline
 $o_i$      &  Agent $i$'s local observation \\ 
 \hline
 $h_i$      &  Agent $i$'s historical representation \\ 
 \hline
 ${c}_{i}^{(0)}$      &  Agent $i$'s initial local embedding \\ 
 \hline
 ${c}_{i}^{(1)}$      &  \makecell[c]{Agent $i$'s local embedding \\after the first communication round}\\ 
 \hline
 ${c}_{i}^{(2)}$      &  \makecell[c]{Agent $i$'s local embedding \\after the second communication round} \\ 
  \hline
 $a_i$      &  Agent $i$'s local action \\ 
 \hline
 $\mathcal{N}_i^{(1)}$      &  The set of agent $i$'s one-hop neighbors\\ 
   \hline
 $\mathcal{\widetilde{N}}_i^{(2)}$       &  The set of agent $i$'s two-hop neighbors\\ 
   \hline
 $\mathcal{N}_i^{(2)}$     &  \makecell[c]{The set of agent $i$'s two-hop neighbors, \\ excluding its one-hop neighbors and itself}\\ 
    \hline
 $T$      &  The controller threshold\\ 
  \hline
 ${z}_{i}$      &  Signal produced by agent $i$'s controller \\ 
  \hline

\end{tabular}
}
\label{table:notations}
\end{table}

\subsection{Two-Hop Communication}
\subsubsection{Communication Protocol}
To enable effective communication and facilitate coordination among agents, AC2C implements a two-round communication mechanism. 

In the first communication round, agent $i$ broadcasts its local embedding $c_i^{(0)}$ to its one-hop neighbors $\mathcal{N}_i^{(1)}$, and it also receives the embeddings from them. Upon receiving messages from $\mathcal{N}_i^{(1)}$ in the first communication round, agent $i$ aggregates the messages as well as its local embedding with an attention-based aggregation module and obtains the updated embedding $c_i^{(1)}$:
\begin{equation}
    c_i^{(1)} = f^{(1)} \left(c_i^{(0)}, c_j^{(0)} \left|\right. j \in  \mathcal{N}_i^{(1)} \right),
\end{equation}
where $f^{(1)}(\cdot,\cdot)$ denotes the first round aggregation function to be introduced in Section \ref{msg_agg}.

Then, agent $i$'s controller leverages $c_i^{(0)}$ and $c_i^{(1)}$ to locally determine whether a second communication round is needed for agent $i$ by outputting a binary signal $z_i$. We will defer the implementation details of the controller to Section \ref{controller}. 

Upon deciding that the second communication round is needed, agent $i$ will inform its one-hop neighbors to initiate the second communication round. In this round, agent $i$ receives messages from its two-hop neighbors $\mathcal{N}_i^{(2)}$, with its one-hop neighbors only acting as relaying nodes. After receiving messages, agent $i$ again aggregates the messages and its local embedding $c_i^{(1)}$ by the aggregation module and obtains the updated embedding $c_i^{(2)}$:
\begin{equation}
    c_i^{(2)} = f^{(2)} \left(c_i^{(1)}, c_j^{(1)} \left|\right. j \in  \mathcal{N}_i^{(2)} \right),
\end{equation}
where $f^{(2)}(\cdot,\cdot)$ denotes the second round aggregation function to be described in Section \ref{msg_agg}.

After two communication rounds, each agent possesses embeddings $c_i^{(0)}$, $c_i^{(1)}$ and $c_i^{(2)}$ (if the second round communication was executed), which will all be concatenated together and fed into the action policy for decision making. 

\subsubsection{Message Aggregation Strategy} \label{msg_agg}
For the message aggregation strategy, we implement an attention module for each communication round. 

In the $n$-th communication round, we first calculate the key $k_i^{(n)}\in\mathbb{R}^{d}$, the query $q_i^{(n)}\in\mathbb{R}^{d}$ and the value $v_i^{(n)}\in\mathbb{R}^{d}$ from $c_i^{(n-1)}$ \citep{Attention,GAT} as:
\begin{equation}
k_i^{(n)}= W_k^{(n)}  c_i^{(n-1)},
\end{equation}
\begin{equation}
q_i^{(n)}= W_q^{(n)}  c_i^{(n-1)},
\end{equation}
\begin{equation}
v_i^{(n)}= W_v^{(n)}  c_i^{(n-1)},
\end{equation}
where $W_k^{(n)}$, $W_q^{(n)}$, $W_v^{(n)}$ are model parameters.

Then, attention weights $\alpha_{ij}^{(n)}$ are obtained with a softmax function:
\begin{equation}
\alpha_{i j}^{(n)}=\operatorname{softmax}\left(e_{i j}^{(n)}\right)=\frac{\exp \left(e_{i j}^{(n)}\right)}{\sum_{k \in \mathcal{N}_{i}^{(n)}} \exp \left(e_{i k}^{(n)}\right)},
\end{equation}
where $e_{ik}^{(n)}=\text{LeakyRelu}(\frac{{q_i^{(n)}}^T{k_k^{(n)}}}{\sqrt{d}})$.

Finally the updated embedding $c_i^{(n)}$ is calculated as:
\begin{align}
c_i^{(n)}
&= f^{(n)} \left(c_i^{(n-1)}, c_j^{(n-1)} \left|\right. j \in  \mathcal{N}_i^{(n)} \right) \\
&=\text{tanh}\left[\sum_{j\in{\mathcal{N}_i^{(n)}}}\alpha_{ij}^{(n)}v_{j}^{(n)}\right].
\end{align}
where $v_{j}^{(1)}$ is the value generated during the first-round communication.
We summarize the communication protocol from the receiver's side in Algorithm \ref{algo_comm_protocol}.

\begin{algorithm*}
\caption{AC2C Communication Protocol at agent $i$}\label{protocol_algorithm}
\begin{algorithmic}[1]
\State \textbf{Inputs:} Initial embedding $c_i^{(0)}$, $c_j^{(0)}$

\State Receive messages from agent $i$'s one-hop neighbors $\mathcal{N}_i^{(1)}$ 
\Comment{The first communication round}
\State Compute the local embedding 
$c_i^{(1)} = f^{(1)} \left(c_i^{(0)}, c_j^{(0)} \left|\right. j \in  \mathcal{N}_i^{(1)} \right)$ in Equation (1)
\State Compute the binary signal 
$z_i = g(c_{i}^{(0)},c_{i}^{(1)};\theta_c, T)$ in Equation (9)
\If{$z_i == 1$}
\State Broadcast a request to initiate the second communication round
\State Receive messages from agent $i$'s two-hop neighbors $\mathcal{N}_i^{(2)}$ with one-hop neighbors acting as relaying nodes \\
\Comment{The second communication round}
\State Compute the local embedding 
$c_i^{(2)} = f^{(2)} \left(c_i^{(1)}, c_j^{(1)} \left|\right. j \in  \mathcal{N}_i^{(2)} \right)$ in Equations (7) and (8)
\Else 
\State Assign $c_i^{(2)} = \bf{0}$ 
\EndIf

\State \textbf{Outputs:} Embedding $c_i^{(0)}$, $c_i^{(1)}$ and $c_i^{(2)}$

\end{algorithmic}
\label{algo_comm_protocol}
\end{algorithm*}

Note that while our method can easily be generalized to its multi-round variations by stacking the communication modules for more complicated applications, we confine it to the two-round case in this paper since we do not observe further performance gain when stacking more than two rounds in our experiments.

\subsection{Two-Hop Controller} \label{controller}


Since the second communication round may induce high communication costs, we propose a local two-hop controller to adaptively prune the unnecessary two-hop communication links to reduce the communication cost. 

During execution, the agent $i$'s controller takes $c_{i}^{(0)}$ and $c_i^{(1)}$ as inputs and generates a signal $z_i$, determining whether to broadcast a request to initiate the second communication round:
\begin{equation} \label{controller_equation}
z_i=\mathds{1}\left[h \left(c_{i}^{(0)},c_{i}^{(1)};\theta_c\right) >T \right]
\end{equation}
where $\theta_c$ is controller's parameters , $z_i \in \{0, 1 \}$ is the binary signal, $T \in (0,1) $ is the threshold, and $\mathds{1} [ \cdot]$ is the indicator function.

We train the controller as a binary classifier in a self-supervised fashion. The training process for this controller is given in Algorithm \ref{training_algorithm}. The loss function for this auxiliary task is formulated as:

\begin{equation}\label{loss_c}
\begin{aligned}
    \mathcal{L}(\theta_{c})=-\mathbb{E}_{\boldsymbol{o}, \boldsymbol{h}} &\left[y_i \log h \left(c_{i}^{(0)},c_{i}^{(1)};\theta_c\right) \right. \\
    & \left.+  (1-y_i) \log \left(1- h \left(c_{i}^{(0)},c_{i}^{(1)};\theta_c\right)\right)\right],
\end{aligned}
\end{equation}
with
\begin{equation}
a^{\Rmnum{1}}_{i}=\pi\left(c_i^{(0)},c_i^{(1)}, \textbf{0} ; \theta_\pi \right),
\end{equation}  
\begin{equation}
a^{\uppercase\expandafter{\romannumeral2}}_{i}=\pi \left(c_i^{(0)}, c_i^{(1)}, c_i^{(2)}; \theta_\pi \right),
\end{equation}

\begin{equation}\label{cls_target}
y_i =\mathds{1}\left[ \lVert a^\Rmnum{1}_{i}- a^{\uppercase\expandafter{\romannumeral2}}_{i} \rVert >T \right],
\end{equation}
where $\pi_{\theta}(\cdot)$ denotes the action policy, $a^{\Rmnum{1}}_{i}$ and $a^{\uppercase\expandafter{\romannumeral2}}_{i}$ denote the logit of agent $i$'s action decisions after receiving messages in the first and the second communication round, respectively.

The underlying idea of Equation \eqref{cls_target} is that once the second-round messages do not contribute much to agent $i$'s action decision, they would be eliminated as redundant information.
Built on this intuition and the training objective given by Equation \eqref{loss_c}, our controller only exploits the second round communication when necessary. In this way, the controller is able to cut off redundant information while maintaining satisfactory performance.

\begin{figure*}[t]
\centering
\subfigure[Traffic Junction]{
\includegraphics[width =  0.22\textwidth]{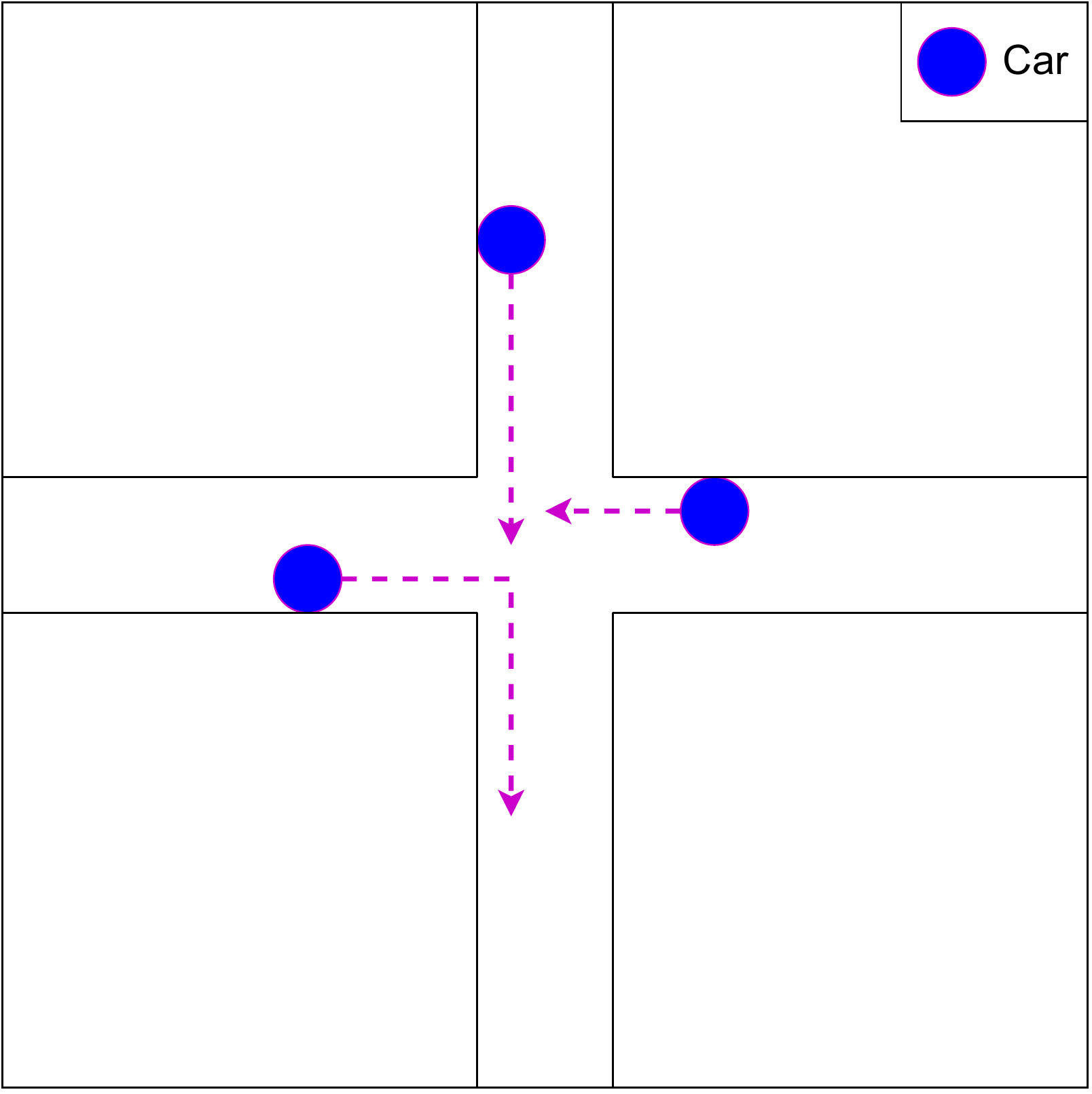}
\label{fig:tj}
}
\subfigure[Cooperative Navigation]{
\includegraphics[width =  0.22\textwidth]{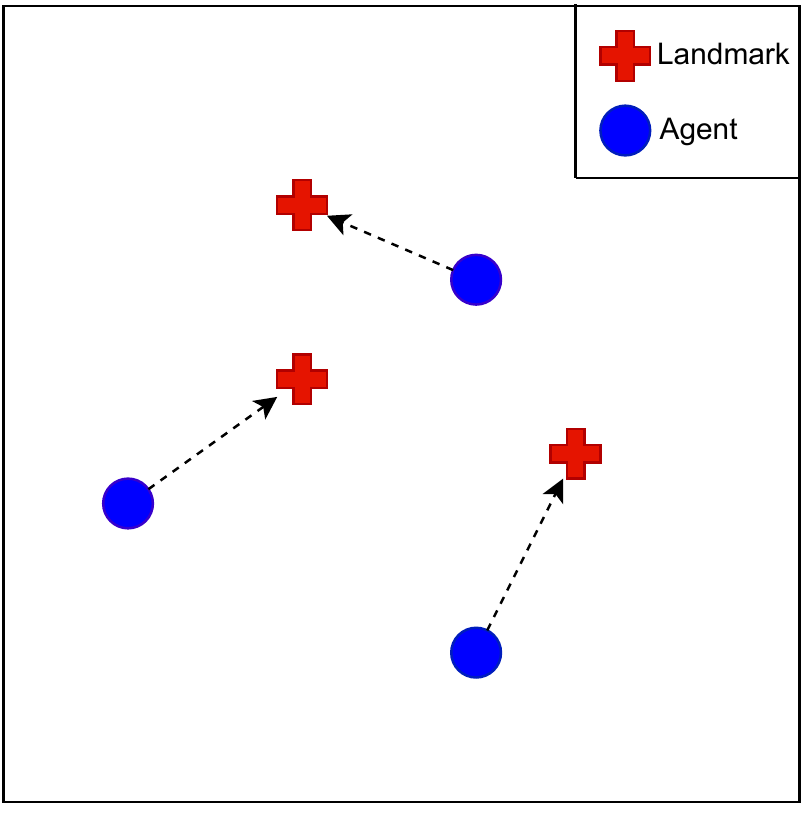}
\label{fig:cn}
}
\subfigure[Predator Prey]{
\includegraphics[width =  0.22\textwidth]{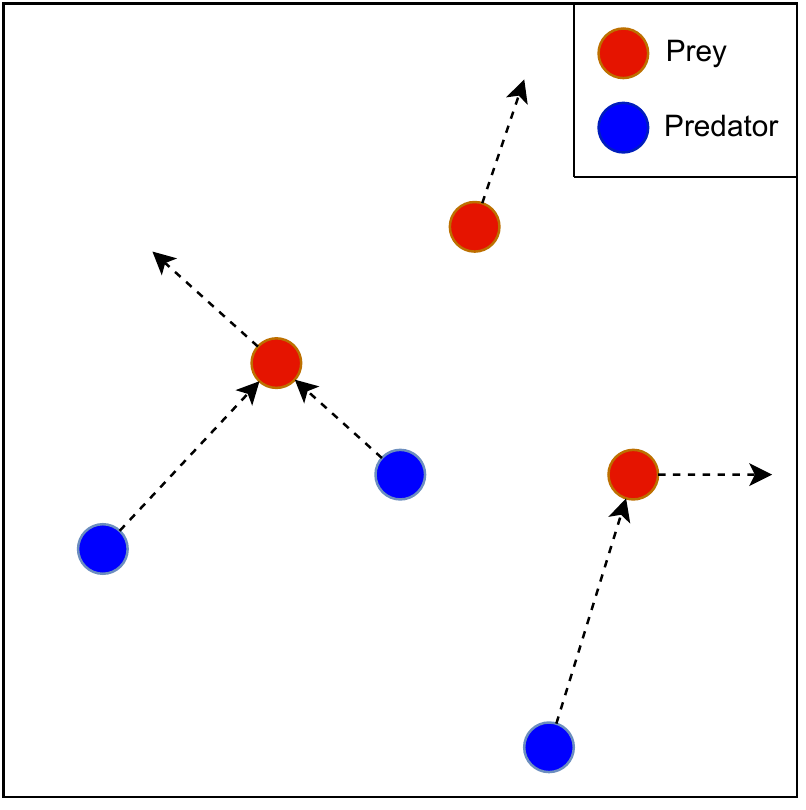}
\label{fig:pp}
}
\caption{Three environments in the experiments: (a) traffic junction, (b) cooperative navigation, and (c) predator prey.
}
\label{fig:environments}
\end{figure*}

\begin{algorithm}[t]

\caption{Training Procedure for the Two-Hop Controller}\label{training_algorithm}
\begin{algorithmic}[1]
\State \textbf{Inputs:} Replay buffer $\mathcal{D}$, the controller threshold $T$
\State \textbf{Initializes:} Controller network parameters $\theta_c$
\State Sample a batch $\mathcal{B}$ with $| \mathcal{B} |$ transitions $(\boldsymbol{o}, \boldsymbol{h}, \boldsymbol{a}, r, \boldsymbol{o}')$ from replay buffer $\mathcal{D}$

\For{$i = 1 \cdots N$}
\State Compute the embeddings $c_i^{(0)}$, $c_i^{(1)}$, $c_i^{(2)}$ in Equations (1), (7) and (8)

\State Compute local action values $a^{\Rmnum{1}}_{i}=\pi_\theta(c_i^{(0)},c_i^{(1)}, \textbf{0})$ and $a^{\uppercase\expandafter{\romannumeral2}}_{i}=\pi_\theta(c_i^{(0)}, c_i^{(1)}, c_i^{(2)})$ in Equations (11) and (12)
        
\State Compute $y_i =\mathds{1}\left[ \lVert a^\Rmnum{1}_{i}- a^{\uppercase\expandafter{\romannumeral2}}_{i} \rVert >T \right]$
in Equation (13)
\State Update $\theta_{c}$ to minimize $\mathcal{L}(\theta_{c})$ in Equation \eqref{loss_c}
\EndFor

\State \textbf{Outputs:} $\theta_c$
\end{algorithmic}
\end{algorithm}

\subsection{Centralized Critic}
We adopt the actor-critic structure that has been wildly used for many single-agent and multi-agent algorithms. Following previous works \citep{COMA, MADDPG}, we leverage a centralized critic network to guide the policy optimization. The critic network shares a similar structure with the actor, but it takes the historical information $\boldsymbol{h}$, observations $\boldsymbol{o}$, and additional predicted actions $\boldsymbol{a}$ from all agents as inputs. A centralized critic network leverages all agents' information to update each agent's gradient. It can greatly alleviate the non-stationary problem. In order to make the implementation scalable, the centralized critic is not needed during execution.

\subsection{Training}
We implement the DDPG and REINFORCE algorithms for different experiments. 

In the DDPG algorithm, we adopt a shared critic with a similar structure to the actor to guide each agent to update its policy under the CTDE paradigm. The centralized critic is updated by the standard TD loss:
\begin{equation}
    \mathcal{L}(\theta_Q) = \mathbb{E}_{\boldsymbol{\tau}, \boldsymbol{h}, \boldsymbol{a}, r, \boldsymbol{\tau}'} \left[\left(y - Q\left(\boldsymbol{\tau}, \boldsymbol{a}; \theta_{Q}\right) \right)^2 \right],
\end{equation}
\begin{equation}
y = r + \gamma Q^{\prime}\left(\boldsymbol{\tau}', \pi^{\prime}\left(\boldsymbol{\tau}'; \theta_{\pi^{\prime}}\right) ; \theta_{Q^{\prime}}\right),
\end{equation}
where $Q^{\prime}$ is the target $Q$ network, $\pi^{\prime}$ is the target actor network, and $\theta_Q$ contains the parameters of the centralized critic network. 
Besides, we update the actor network parameters $\theta_\pi$ by the sampled policy gradient:
\begin{equation}
\nabla_{\theta_\pi}J(\theta_\pi) = \mathbb{E}_{\boldsymbol{\tau}, \boldsymbol{h}, \boldsymbol{a}, r, \boldsymbol{\tau}'}  \left[\nabla_{\theta_\pi}\pi (\tau_i ; \theta_{\pi})\nabla_{a}Q(\tau_i,a_i ; \theta_{Q})|_{a_{i}=\pi(\tau_i)} \right]
.
\end{equation}

In the traffic junction experiments, we adopt the REINFORCE algorithm with baseline \citep{sutton1999policy} to learn the actor policy. We update the policy network parameters $\theta_\mu$ by the following equation:
\begin{equation}
\nabla_{\theta_\pi}J(\theta_\pi) = \mathbb{E}_{\boldsymbol{\tau}, \boldsymbol{h}, \boldsymbol{a}, r, \boldsymbol{\tau}'} \left[ \left( G-b(\boldsymbol{\tau})\right) \nabla_{\theta_\pi} \log \pi\left(\tau_i; \theta_\pi \right)  \right],
\end{equation}

where $G$ is the episodic return and $b(\cdot)$ is the counterfactual baseline.
In order to accelerate the training, we share the feature encoder and action policy parameters across agents.

\section{Experiments}

\begin{figure}[t]
\centering
\subfigure[Cooperative navigation]{
\includegraphics[width=0.32\textwidth]{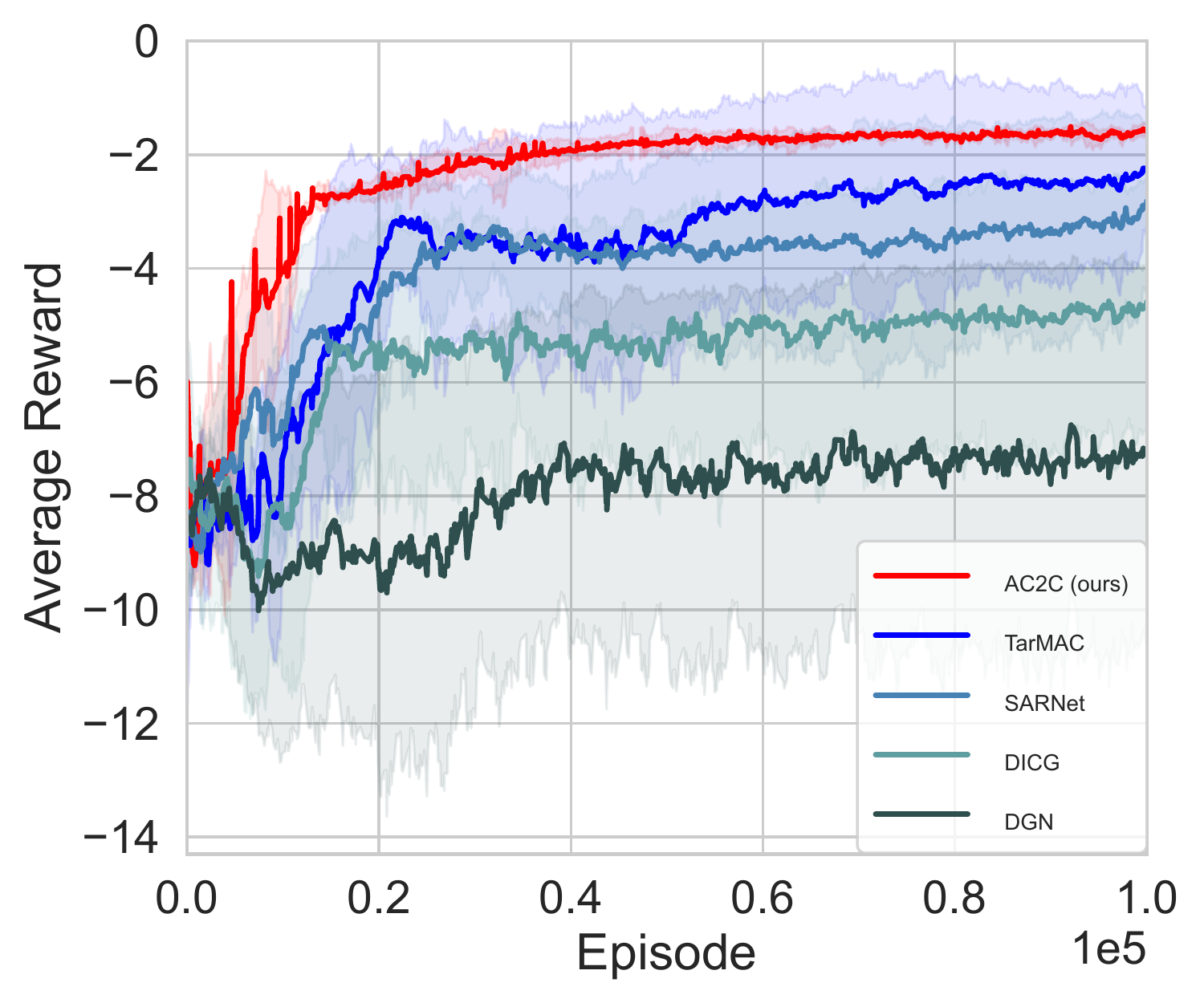}
}
\vfill
\subfigure[Predator prey]{
\includegraphics[width=0.32\textwidth]{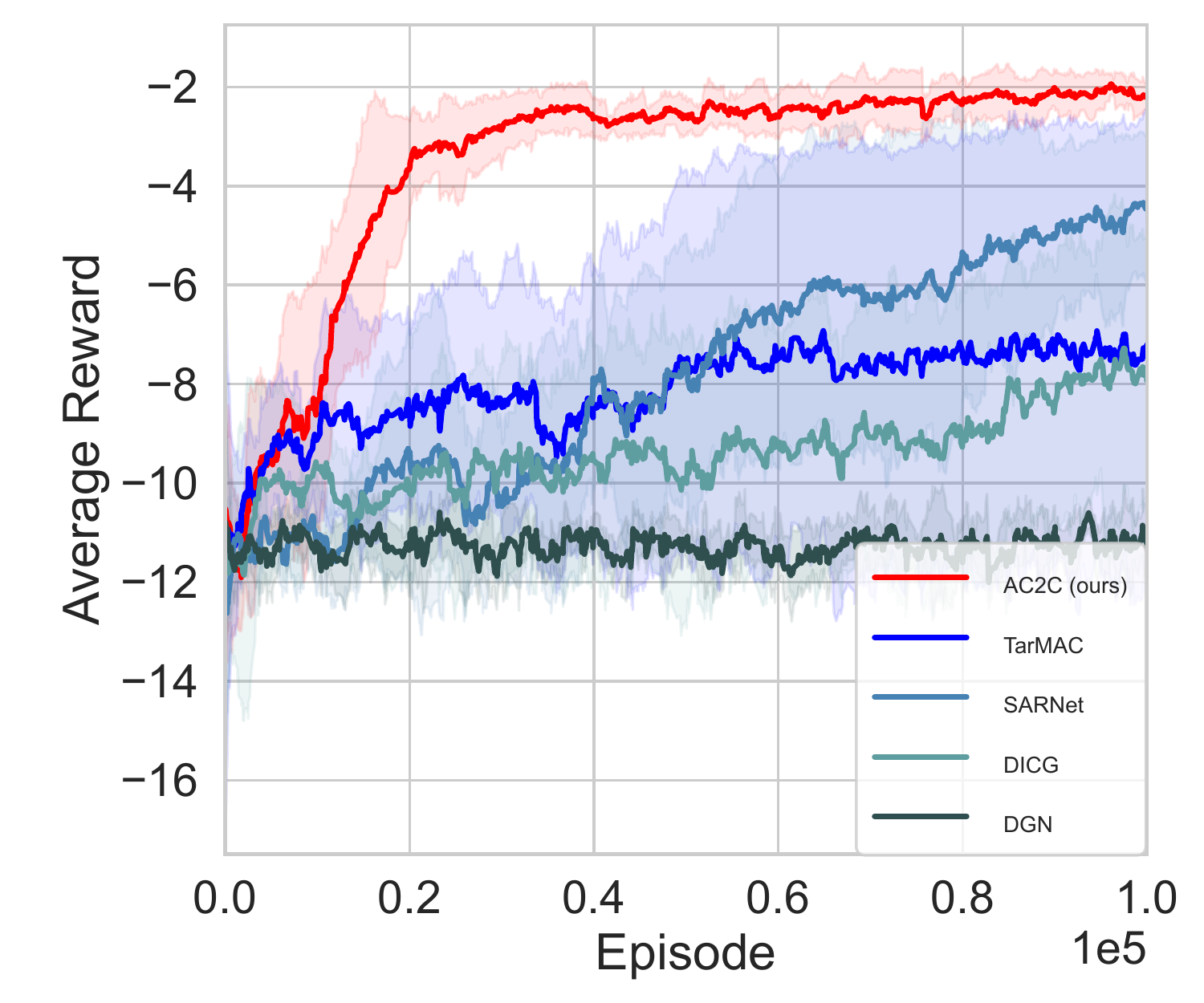}
}
\caption{Training curves of cooperative navigation and predator prey.}
\label{fig:PP_CN_training}
\end{figure}

We evaluate the proposed AC2C in three environments, namely, traffic junction, cooperative navigation and predator prey, as illustrated in Fig \ref{fig:environments}.
Following \citep{i2c,ic3}, our method as well as the baseline methods are implemented on top of the REINFORCE algorithm \citep{sutton1999policy} in the traffic junction environment, and on top of the DDPG \citep{DDPG} in the predator prey and cooperative navigation environments.

All presented results are average performance over five random seeds. The shaded area in each figure is the standard deviation. 
\subsection{Baselines}
In this work, we compare our method with baselines including TarMAC \citep{Tarmac}, SARNet \citep{sarnet}, DICG \citep{dicg} and DGN \citep{DGN}. TarMAC achieves multi-round communication with a back-and-forth method \citep{Tarmac}. SARNet leverages a memory-based mechanism to solve cooperative multi-agent tasks. DICG and DGN are both typical GNN-based methods. DICG utilizes GCN layers and attention mechanisms to accomplish message aggregation, while DGN considers dynamic communication graphs and adopts GCN layers to conduct message aggregation.


In our AC2C communication protocol, each agent exchanges local information with its one-hop neighbors in the first communication round and exchanges messages with its two-hop neighbors in the second communication round. We set a GNN-based protocol for all baselines. Specifically, each agent communicates with its one-hop neighbors in the first communication round. Once all the agents have aggregated the information of one-hop neighbors, they will communicate again with one-hop neighbors in the next round to obtain information from further-away agents. 
To quantify the communication cost, we calculate the number of active communication links, where $w$ bits messages are transmitted through each link. The cost of the first communication round $\text{Cost}^{(1)}$ for the GNN-based method and AC2C is calculated as:
\begin{equation}
\text{Cost}^{(1)}=\sum_{i} | \mathcal{N}_i^{(1)} | \cdot w
\end{equation}
where the $| \mathcal{N}_i^{(1)} |$ is the number of agent $i$'s one-hop neighbors.
The cost of the second communication round $\text{Cost}^{(2)}$ is computed as:
\begin{equation}
\text{Cost}^{(2)}=
\begin{cases}
\sum_{i}{| \mathcal{N}_i^{(2)} |}\cdot 2w,& \text{AC2C}\\
\sum_i{| \mathcal{N}_i^{(1)} |}\cdot w&, \text{GNN-based methods}
\end{cases}
\end{equation}
As the AC2C protocol transmits two-hop messages through a relaying node, the communication cost in the second round should be doubled.

\subsection{Environments} 
\subsubsection{Traffic Junction}
The simulated traffic junction environment introduced by \cite{CommNet}, consists of 20 cars moving along predefined routes with one or more road junctions, as shown in Figure \ref{fig:tj}.
The goal for each car is to arrive at the destination while avoiding collisions with other cars.
Following \cite{i2c,ic3}, each car's field of view is set to 0, but it can communicate with other cars within its communication range.
Each car has only two actions: brake or gas (move forward).
The reward for each car includes a linear time penalty $-0.01\tau$, where $\tau$ is the number of timesteps after a car becomes active, and a penalty of -20 induced by collisions.
In the experiments, we consider two modes of traffic junction: a medium mode and a hard mode.
Particularly, the dimension of the map is set to $6\times6$, and the number of intersections is set as 1 in the medium mode. In the hard mode, the dimension of the map is $9\times9$, and the number of intersections is set as 4. In order to effectively compare the performance, we evaluate the success rate under 20000 testing episodes. We regard an episode as successful if no collision happens during this episode. 

\subsubsection{Cooperative Navigation.} 

The goal of cooperative navigation \cite{MADDPG} is for several agents to cover landmarks respectively, as shown in Figure \ref{fig:cn}.
In our experiments, 10 agents try to occupy 10 fixed landmarks, where each agent obtains partial observation of the environment. Specifically, an agent only knows its own position as well as velocity, and the positions of all the landmarks. 
In this environment, each agent will get a bonus when it approaches the landmark but will receive a penalty when it collides with other agents. All agent are initialized at random positions in every episode. The episode length is set as 50 timesteps. We use the average reward per timestep of each agent as the evaluation metric.

\subsubsection{Predator Prey}

Following previous work \citep{i2c, sarnet}, the predator prey environment is a cooperative multi-agent task as is shown in Figure \ref{fig:pp}.
The goal is for the predators to capture as many preys as possible during a given time period.
The observations of each predator include its position, velocity, the two closest preys' positions, and two closest predators' positions. 
As the preys move slightly faster than the predators, the predators need to learn how to capture the preys cooperatively.
We generate an environment with 10 agents (predators) and 10 preys, where the actions of the preys are controlled by the bots in I2C\citep{i2c}.
Each predator gets a bonus when it captures a prey while receiving a penalty when a collision among predators happens. The evaluation metric is set the same as cooperative navigation.

\begin{table}[t]
	\centering
	\caption{Success rates and communication overhead per timestep of traffic junction.}
	\resizebox{0.465\textwidth}{!}{
	\begin{tabular}{c|c|c|c|c}
        \hline
        &  \multicolumn{4}{c}{Traffic junction}  \\ \cline{2-5}
        \quad& \multicolumn{2}{c}{Medium mode}& \multicolumn{2}{c}{Hard mode}\\
        \cline{2-5}
        \quad& Success rate& \begin{tabular}[c]{@{}c@{}} Communication\\  overhead $(10^5 \, \text{bits})$\end{tabular}   & Success rate& \begin{tabular}[c]{@{}c@{}} Communication\\  overhead $(10^5 \, \text{bits})$ \end{tabular} \\
        \hline
        AC2C (ours) & \bf{95.33$\pm$0.21} & \bf{2.972$\pm$0.636} & \bf{71.85$\pm$1.45}&\bf{5.030$\pm$0.701}\\
        TarMAC&93.81$\pm$0.17 & 3.790$\pm$0.245 & 49.23$\pm$1.06&6.979$\pm$0.705\\
        SarNet&92.16$\pm$0.97 & 3.623$\pm$0.807 & 46.73$\pm$1.31&7.063$\pm$1.029\\
        DICG&95.21$\pm$1.12 & 4.003$\pm$0.794 & 53.71$\pm$2.77&7.613$\pm$1.217\\
        DGN&86.37$\pm$1.27 & 3.918$\pm$0.930 & 17.43$\pm$3.91&7.077$\pm$0.590\\
        \hline
    \end{tabular}
    }
    \label{table: TJ_tradeoff}
\end{table}

\begin{table}[t]
	\centering
	\caption{Performance and communication cost per timestep of cooperative navigation and predator prey.}
	\resizebox{0.475\textwidth}{!}{
	\begin{tabular}{c|c|c|c|c}
        \hline
        \quad& \multicolumn{2}{c|}{Cooperative navigation}& \multicolumn{2}{c}{Predator prey}\\
        \hline
        \quad& Reward & \begin{tabular}[c]{@{}c@{}} Communication\\  overhead $(10^5 \, \text{bits})$\end{tabular}   & Reward & \begin{tabular}[c]{@{}c@{}} Communication\\  overhead $(10^5 \, \text{bits})$\end{tabular} \\
        \hline
        AC2C (ours) & \bf{-1.573$\pm$0.221} & \bf{5.636$\pm$0.614} & \bf{-2.502$\pm$0.402} & \bf{4.878$\pm$0.520}\\
        TarMAC & -2.043$\pm$0.216 & 8.526$\pm$0.934 & -3.654$\pm$0.679 & 7.780$\pm$1.354\\
        SarNet & -3.055$\pm$0.276 & 9.362$\pm$1.106 & -4.042$\pm$0.248 & 6.344$\pm$1.630\\
        DICG & -5.538$\pm$0.378  & 7.770$\pm$1.484 & -5.478$\pm$0.438 & 7.046$\pm$2.108\\
        DGN& -6.696$\pm$0.817  &7.688$\pm$1.496 & -9.250$\pm$0.956 &6.664$\pm$2.318\\
        \hline
    \end{tabular}
    }
    \label{table: CN_PP_tradeoff}
\end{table}

\subsection{Results}

We first investigate the trade-off between the performance and the communication overhead. As illustrated in Table \ref{table: TJ_tradeoff}, Table \ref{table: CN_PP_tradeoff} and Figure \ref{fig:PP_CN_training}, for all three environments, our proposed AC2C significantly outperforms all baselines while maintaining the lowest communication overhead.
Detailed observations are elaborated below for each environment.
\begin{table*}[!htbp]
	\centering
	\caption{Average test rewards and communication cost per timestep of each agent under different communication range in cooperative navigation.}
        \resizebox{1\textwidth}{!}{
	\begin{tabular}{c|c|c|c|c|c|c|c|c|c|c}
        \hline
        \quad & \multicolumn{2}{c|}{AC2C (ours)} &\multicolumn{2}{c|}{TarMAC} &\multicolumn{2}{c|}{SARNet} &\multicolumn{2}{c|}{DICG} &\multicolumn{2}{c}{DGN} \\
        \hline
        \quad & Reward & \begin{tabular}[c]{@{}c@{}}Communication\\ overhead $(10^5 \, \text{bits})$\end{tabular} & Reward & \begin{tabular}[c]{@{}c@{}}Communication\\ overhead $(10^5 \, \text{bits})$\end{tabular} & Reward & \begin{tabular}[c]{@{}c@{}}Communication\\ overhead $(10^5 \, \text{bits})$\end{tabular} & Reward & \begin{tabular}[c]{@{}c@{}}Communication\\  $(10^5 \, \text{bits})$\end{tabular} & Reward & \begin{tabular}[c]{@{}c@{}}Communication\\ overhead $(10^5 \, \text{bits})$\end{tabular}\\
        \hline
        0.3 & \bf{-1.841}$\pm$0.221 & \bf{0.152$\pm$0.030}&  -3.588$\pm$0.353 & 0.198$\pm$0.022&
        -5.731$\pm$0.884 & 0.166$\pm$0.030 & -6.259$\pm$0.575 &0.194$\pm$0.028 &-10.04$\pm$1.025& 0.198$\pm$0.024\\
        0.5& \bf{-1.623$\pm$0.357}  & \bf{1.084$\pm$0.256} & -3.933$\pm$0.423 & 1.262$\pm$0.310 & -5.063$\pm$0.371 & 1.396$\pm$0.364 & -6.629$\pm$0.593 & 1.504$\pm$0.444& -8.166$\pm$0.732 & 1.424$\pm$0.334 \\
        1.0& \bf{-1.573$\pm$0.221}  &\bf{5.636$\pm$0.614}& -2.043$\pm$0.216 & 8.526$\pm$0.934& -3.055$\pm$0.276 &9.362$\pm$1.106& -5.538$\pm$0.378 & 7.770$\pm$1.484 &-6.696$\pm$0.817 & 7.688$\pm$1.496\\
        1.5& \bf{-1.267$\pm$0.158}  &\bf{7.132$\pm$0.652}& -1.418$\pm$0.034 &9.700$\pm$1.814& -1.500$\pm$0.054 &11.47$\pm$1.198& 1.700$\pm$0.344 &8.492$\pm$1.036& -4.603$\pm$0.551&9.634$\pm$0.924\\
        \hline
    \end{tabular}
    }
    \label{table: CN_comm_range}
\end{table*}

\begin{table*}[h]
	\centering
	\caption{Average test rewards and communication cost per timestep of each agent under different communication range in predator prey.}
        \resizebox{1\textwidth}{!}{
	\begin{tabular}{c|c|c|c|c|c|c|c|c|c|c}
        \hline
        \quad & \multicolumn{2}{c|}{AC2C (ours)} &\multicolumn{2}{c|}{TarMAC} &\multicolumn{2}{c|}{SARNet} &\multicolumn{2}{c|}{DICG} &\multicolumn{2}{c}{DGN} \\
        \hline
        \quad & Reward & \begin{tabular}[c]{@{}c@{}}Communication\\ overhead $(10^5 \, \text{bits})$\end{tabular} & Reward & \begin{tabular}[c]{@{}c@{}}Communication\\ overhead $(10^5 \, \text{bits})$\end{tabular} & Reward & \begin{tabular}[c]{@{}c@{}}Communication\\ overhead $(10^5 \, \text{bits})$\end{tabular} & Reward & \begin{tabular}[c]{@{}c@{}}Communication\\  $(10^5 \, \text{bits})$\end{tabular} & Reward & \begin{tabular}[c]{@{}c@{}}Communication\\ overhead $(10^5 \, \text{bits})$\end{tabular}\\
        \hline
        0.3 & \bf{-5.443}$\pm$0.447 & \bf{0.166$\pm$0.014}&  -7.504$\pm$1.080 & 0.188$\pm$0.016 &
        -7.082$\pm$492.3 & 0.174$\pm$0.026 & -7.130$\pm$0.952 & 0.19$\pm$0.022 &-10.44$\pm$0.849& 0.178$\pm$0.018\\
        0.5& \bf{-3.673$\pm$0.390}  & \bf{0.784$\pm$0.062} & -5.714$\pm$0.651 & 1.388$\pm$0.192 & -5.343$\pm$0.404 & 1.346$\pm$0.382 & -6.026$\pm$0.624 & 1.428$\pm$0.634& -9.566$\pm$1.162 & 1.522$\pm$0.410 \\
        1.0& \bf{-2.502$\pm$0.402}  &\bf{4.878$\pm$0.520}& -3.654$\pm$0.679 & 7.780$\pm$1.354& -4.042$\pm$0.248 &6.344$\pm$0.630& -5.478$\pm$0.438 & 7.046$\pm$2.108 &-9.250$\pm$0.956&6.664$\pm$2.318\\
        1.5& \bf{-2.034$\pm$0.211}  &\bf{6.944$\pm$0.638}& -3.254$\pm$0.679 &8.586$\pm$1.024& -2.876$\pm$0.341 &7.370$\pm$0.486& 4.100$\pm$0.294 &8.194$\pm$0.944& -5.487$\pm$0.951&9.246$\pm$0.722\\
        \hline
    \end{tabular}
    }
    \label{table: PP_comm_range}
\end{table*}
In the traffic junction environment, we see from Table \ref{table: TJ_tradeoff} that in the medium mode, AC2C performs slightly better than other baselines with a lower communication overhead. We observe that, very few communication links are constructed since the cars are scattered sparsely across the map. Although agents sometimes obtain information of more distant agents, most of the time, they can only utilize the information of their one-hop neighbors. In the hard mode, AC2C demonstrates a substantial performance gain compared with all the baselines, in both the reward and communication overhead. It illustrates the importance of the communication mechanism design in such difficult environments. 
For cooperative navigation and predator prey tasks, AC2C consistently outperforms the baselines. This is attributed to the long-range information exchange enabled by our protocol. We discover that information belonging to farther agents can effectively improve the current agent's action decisions, e.g., this information can point out where the landmarks are located. In addition, the controller effectively prunes irrelevant messages to make AC2C agents maintain the lowest communication cost. 

We also examine the influence of the communication range $L$ on the performance.
In particular, we set the communication range as $L = 0.3, 0.5, 1.0, 1.5$ in the cooperative navigation and predator prey tasks. As illustrated in Tables \ref{table: CN_comm_range} and \ref{table: PP_comm_range}, AC2C consistently outperforms all the baselines and achieves low communication overhead.
For all the methods, as the communication range shrinks, both the rewards and communication overhead are decreased, which is due to the smaller number of one-hop and two-hop neighbors.
The results demonstrate that the controller in our method effectively identifies and prunes the irrelevant communication links without incurring performance degradation.

\begin{figure}[t] 
\centering 
\includegraphics[width=5.8cm]{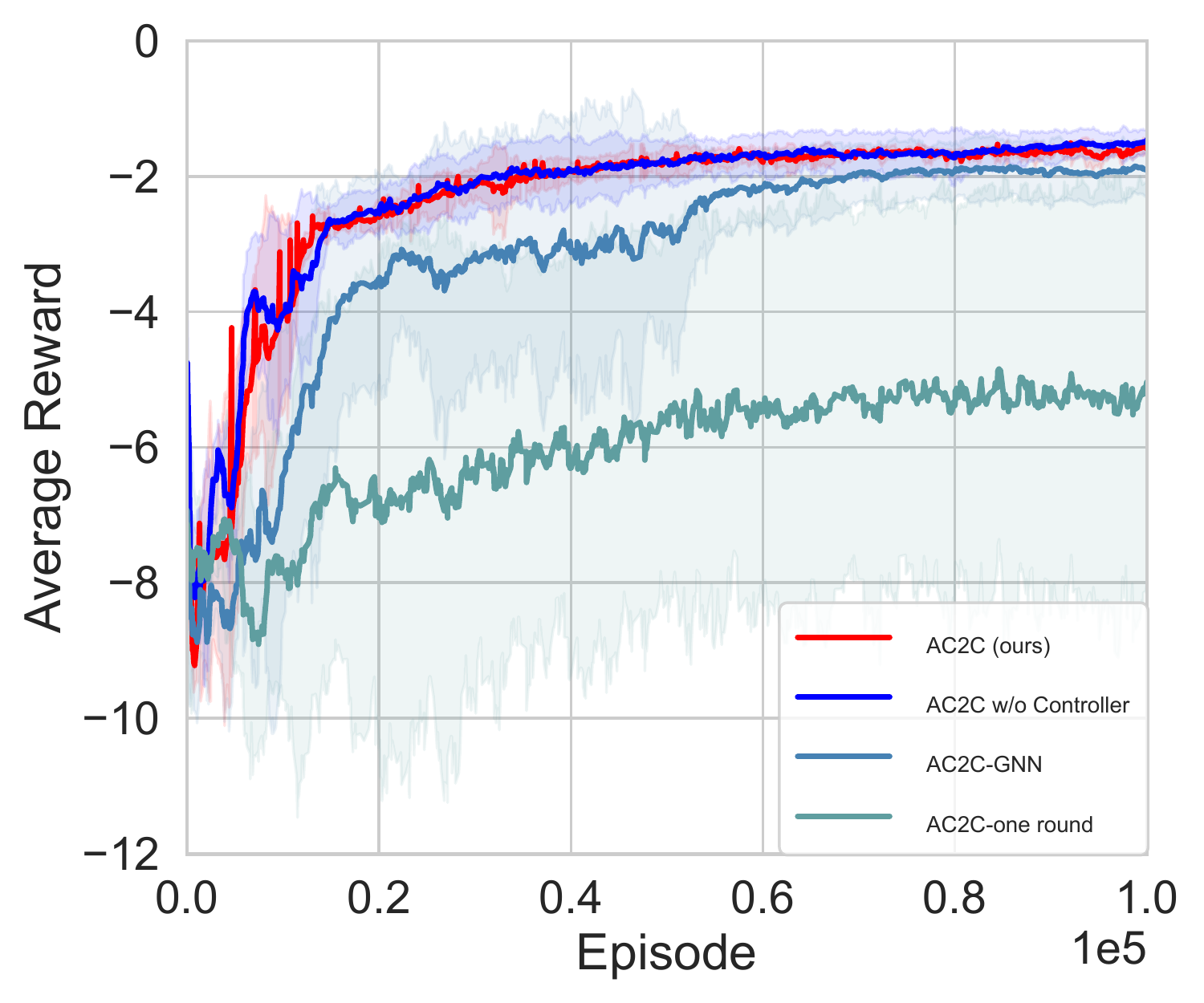} 
\caption{The performance between different AC2C versions.
} 
\label{difference_AC2C} 
\end{figure}

\begin{table}[ht]
\tiny
	\centering
	\caption{The impact of different threshold $T$ in cooperative navigation.}
	\resizebox{0.45\textwidth}{!}{
	\begin{tabular}{c|c|c}
        \hline
        \quad & Reward & Communication Overhead ($10^5$ bits) \\
        \hline
        $T=0$ & -1.579$\pm$0.195 & 9.546$\pm$0.878 \\
        \hline
        $T=0.1$ & -1.546$\pm$0.100 & 8.650$\pm$0.524 \\
        \hline
        $T=0.2$ & -1.539$\pm$0.195 & 8.072$\pm$0.622 \\
        \hline
        $T=0.3$ & -1.565$\pm$0.207& 6.672$\pm$0.724 \\
        \hline
        $T=0.4$ & -1.553$\pm$0.199 & 6.362$\pm$0.842 \\
        \hline
        $T=0.5$ & -1.573$\pm$0.221 & 5.636$\pm$0.614 \\
        \hline
        $T=0.6$ & -3.855$\pm$0.648 & 4.462 $\pm$0.686\\
        \hline
    \end{tabular}
    }
    \label{table: different_threshold}
\end{table}
\textbf{Ablation Study}. In order to demonstrate the effectiveness of the AC2C protocol, we conduct the ablation study on AC2C under the cooperative navigation environment. AC2C w/o controller refers to AC2C without the controller, where the second-round communication always happens; AC2C-GNN adopts the GNN-based protocol to communicate rather than ours; and AC2C one round indicates that the second-round communication never happens. As shown in Figure \ref{difference_AC2C}, AC2C achieves significant performance gains compared to AC2C-GNN and AC2C-one round. Moreover, it shows that the controller can effectively reduce the communication overhead without performance degradation. We next test the impact of the threshold $T$ on the controller in Table \ref{table: different_threshold}. When the value of $T$ varies from 0 to 0.5, the communication cost reduces, but the performance does not drop significantly. It illustrates that the controller helps to prune the irrelevant information. However, the performance dramatically deteriorates when the threshold reaches 0.6, where the controller cannot retain enough valuable information. In this case, the overall performance is similar to that of AC2C-one round. In other reported results, we did a grid search to find the optimal $T$.

\section{Conclusion}
In this paper, we introduce an effective communication protocol for cooperative multi-agent reinforcement learning systems, which helps agents to obtain valuable messages from agents outside their communication range. A communication controller is introduced to reduce the communication overhead while maintaining performance. Extensive experiments show that the proposed method outperforms all the baselines regarding both the reward and communication overhead in the three considered environments. This study illustrates the importance of developing adaptive multi-hop communication protocols for multi-agent reinforcement learning systems.

\section{Acknowledgment}
This work was supported by the NSFC/RGC Collaborative Research Scheme (Project No. CRS\_HKUST603/22).


\bibliographystyle{ACM-Reference-Format} 

\balance

\bibliography{aamas23.bib}


\begin{thebibliography}{31}


\ifx \showCODEN    \undefined \def \showCODEN     #1{\unskip}     \fi
\ifx \showDOI      \undefined \def \showDOI       #1{#1}\fi
\ifx \showISBNx    \undefined \def \showISBNx     #1{\unskip}     \fi
\ifx \showISBNxiii \undefined \def \showISBNxiii  #1{\unskip}     \fi
\ifx \showISSN     \undefined \def \showISSN      #1{\unskip}     \fi
\ifx \showLCCN     \undefined \def \showLCCN      #1{\unskip}     \fi
\ifx \shownote     \undefined \def \shownote      #1{#1}          \fi
\ifx \showarticletitle \undefined \def \showarticletitle #1{#1}   \fi
\ifx \showURL      \undefined \def \showURL       {\relax}        \fi
\providecommand\bibfield[2]{#2}
\providecommand\bibinfo[2]{#2}
\providecommand\natexlab[1]{#1}
\providecommand\showeprint[2][]{arXiv:#2}

\bibitem[\protect\citeauthoryear{Busoniu, Babuska, and De~Schutter}{Busoniu
  et~al\mbox{.}}{2008}]%
        {busoniu2008comprehensive}
\bibfield{author}{\bibinfo{person}{Lucian Busoniu}, \bibinfo{person}{Robert
  Babuska}, {and} \bibinfo{person}{Bart De~Schutter}.}
  \bibinfo{year}{2008}\natexlab{}.
\newblock \showarticletitle{A comprehensive survey of multiagent reinforcement
  learning}.
\newblock \bibinfo{journal}{\emph{IEEE Transactions on Systems, Man, and
  Cybernetics, Part C (Applications and Reviews)}} \bibinfo{volume}{38},
  \bibinfo{number}{2} (\bibinfo{year}{2008}), \bibinfo{pages}{156--172}.
\newblock


\bibitem[\protect\citeauthoryear{Claus and Boutilier}{Claus and
  Boutilier}{1998}]%
        {CTCE}
\bibfield{author}{\bibinfo{person}{Caroline Claus} {and} \bibinfo{person}{Craig
  Boutilier}.} \bibinfo{year}{1998}\natexlab{}.
\newblock \showarticletitle{The dynamics of reinforcement learning in
  cooperative multiagent systems}.
\newblock \bibinfo{journal}{\emph{AAAI/IAAI}} \bibinfo{volume}{1998},
  \bibinfo{number}{746-752} (\bibinfo{year}{1998}), \bibinfo{pages}{2}.
\newblock


\bibitem[\protect\citeauthoryear{Das, Gervet, Romoff, Batra, Parikh, Rabbat,
  and Pineau}{Das et~al\mbox{.}}{2019}]%
        {Tarmac}
\bibfield{author}{\bibinfo{person}{Abhishek Das},
  \bibinfo{person}{Th{\'e}ophile Gervet}, \bibinfo{person}{Joshua Romoff},
  \bibinfo{person}{Dhruv Batra}, \bibinfo{person}{Devi Parikh},
  \bibinfo{person}{Mike Rabbat}, {and} \bibinfo{person}{Joelle Pineau}.}
  \bibinfo{year}{2019}\natexlab{}.
\newblock \showarticletitle{Tarmac: Targeted multi-agent communication}. In
  \bibinfo{booktitle}{\emph{International Conference on Machine Learning}}.
  PMLR, \bibinfo{pages}{1538--1546}.
\newblock


\bibitem[\protect\citeauthoryear{Ding, Huang, and Lu}{Ding
  et~al\mbox{.}}{2020}]%
        {i2c}
\bibfield{author}{\bibinfo{person}{Ziluo Ding}, \bibinfo{person}{Tiejun Huang},
  {and} \bibinfo{person}{Zongqing Lu}.} \bibinfo{year}{2020}\natexlab{}.
\newblock \showarticletitle{Learning individually inferred communication for
  multi-agent cooperation}.
\newblock \bibinfo{journal}{\emph{Advances in Neural Information Processing
  Systems}}  \bibinfo{volume}{33} (\bibinfo{year}{2020}),
  \bibinfo{pages}{22069--22079}.
\newblock


\bibitem[\protect\citeauthoryear{Du, Liu, Moens, Liu, Ren, Wang, Chen, and
  Zhang}{Du et~al\mbox{.}}{2021}]%
        {corr_commu}
\bibfield{author}{\bibinfo{person}{Yali Du}, \bibinfo{person}{Bo Liu},
  \bibinfo{person}{Vincent Moens}, \bibinfo{person}{Ziqi Liu},
  \bibinfo{person}{Zhicheng Ren}, \bibinfo{person}{Jun Wang},
  \bibinfo{person}{Xu Chen}, {and} \bibinfo{person}{Haifeng Zhang}.}
  \bibinfo{year}{2021}\natexlab{}.
\newblock \showarticletitle{Learning correlated communication topology in
  multi-agent reinforcement learning}. In \bibinfo{booktitle}{\emph{Proceedings
  of the 20th International Conference on Autonomous Agents and MultiAgent
  Systems}}. \bibinfo{pages}{456--464}.
\newblock


\bibitem[\protect\citeauthoryear{Foerster, Assael, De~Freitas, and
  Whiteson}{Foerster et~al\mbox{.}}{2016}]%
        {DIAL}
\bibfield{author}{\bibinfo{person}{Jakob Foerster},
  \bibinfo{person}{Ioannis~Alexandros Assael}, \bibinfo{person}{Nando
  De~Freitas}, {and} \bibinfo{person}{Shimon Whiteson}.}
  \bibinfo{year}{2016}\natexlab{}.
\newblock \showarticletitle{Learning to communicate with deep multi-agent
  reinforcement learning}.
\newblock \bibinfo{journal}{\emph{Advances in neural information processing
  systems}}  \bibinfo{volume}{29} (\bibinfo{year}{2016}).
\newblock


\bibitem[\protect\citeauthoryear{Foerster, Farquhar, Afouras, Nardelli, and
  Whiteson}{Foerster et~al\mbox{.}}{2018}]%
        {COMA}
\bibfield{author}{\bibinfo{person}{Jakob Foerster}, \bibinfo{person}{Gregory
  Farquhar}, \bibinfo{person}{Triantafyllos Afouras}, \bibinfo{person}{Nantas
  Nardelli}, {and} \bibinfo{person}{Shimon Whiteson}.}
  \bibinfo{year}{2018}\natexlab{}.
\newblock \showarticletitle{Counterfactual multi-agent policy gradients}. In
  \bibinfo{booktitle}{\emph{Proceedings of the AAAI conference on artificial
  intelligence}}, Vol.~\bibinfo{volume}{32}.
\newblock


\bibitem[\protect\citeauthoryear{Gupta, Egorov, and Kochenderfer}{Gupta
  et~al\mbox{.}}{2017}]%
        {gupta2017cooperative}
\bibfield{author}{\bibinfo{person}{Jayesh~K Gupta}, \bibinfo{person}{Maxim
  Egorov}, {and} \bibinfo{person}{Mykel Kochenderfer}.}
  \bibinfo{year}{2017}\natexlab{}.
\newblock \showarticletitle{Cooperative multi-agent control using deep
  reinforcement learning}. In \bibinfo{booktitle}{\emph{International
  conference on autonomous agents and multiagent systems}}. Springer,
  \bibinfo{pages}{66--83}.
\newblock


\bibitem[\protect\citeauthoryear{Jiang, Dun, Huang, and Lu}{Jiang
  et~al\mbox{.}}{2019}]%
        {DGN}
\bibfield{author}{\bibinfo{person}{Jiechuan Jiang}, \bibinfo{person}{Chen Dun},
  \bibinfo{person}{Tiejun Huang}, {and} \bibinfo{person}{Zongqing Lu}.}
  \bibinfo{year}{2019}\natexlab{}.
\newblock \showarticletitle{Graph Convolutional Reinforcement Learning}. In
  \bibinfo{booktitle}{\emph{International Conference on Learning
  Representations}}.
\newblock


\bibitem[\protect\citeauthoryear{Jiang and Lu}{Jiang and Lu}{2018}]%
        {atoc}
\bibfield{author}{\bibinfo{person}{Jiechuan Jiang} {and}
  \bibinfo{person}{Zongqing Lu}.} \bibinfo{year}{2018}\natexlab{}.
\newblock \showarticletitle{Learning attentional communication for multi-agent
  cooperation}.
\newblock \bibinfo{journal}{\emph{Advances in neural information processing
  systems}}  \bibinfo{volume}{31} (\bibinfo{year}{2018}).
\newblock


\bibitem[\protect\citeauthoryear{Kingma and Ba}{Kingma and Ba}{2014}]%
        {adam}
\bibfield{author}{\bibinfo{person}{Diederik~P Kingma} {and}
  \bibinfo{person}{Jimmy Ba}.} \bibinfo{year}{2014}\natexlab{}.
\newblock \showarticletitle{Adam: A method for stochastic optimization}.
\newblock \bibinfo{journal}{\emph{arXiv preprint arXiv:1412.6980}}
  (\bibinfo{year}{2014}).
\newblock


\bibitem[\protect\citeauthoryear{Li, Gupta, Morales, Allen, and
  Kochenderfer}{Li et~al\mbox{.}}{2021}]%
        {dicg}
\bibfield{author}{\bibinfo{person}{Sheng Li}, \bibinfo{person}{Jayesh~K Gupta},
  \bibinfo{person}{Peter Morales}, \bibinfo{person}{Ross Allen}, {and}
  \bibinfo{person}{Mykel~J Kochenderfer}.} \bibinfo{year}{2021}\natexlab{}.
\newblock \showarticletitle{Deep Implicit Coordination Graphs for Multi-agent
  Reinforcement Learning}. In \bibinfo{booktitle}{\emph{Proceedings of the 20th
  International Conference on Autonomous Agents and MultiAgent Systems}}.
  \bibinfo{pages}{764--772}.
\newblock


\bibitem[\protect\citeauthoryear{Lillicrap, Hunt, Pritzel, Heess, Erez, Tassa,
  Silver, and Wierstra}{Lillicrap et~al\mbox{.}}{2015}]%
        {DDPG}
\bibfield{author}{\bibinfo{person}{Timothy~P Lillicrap},
  \bibinfo{person}{Jonathan~J Hunt}, \bibinfo{person}{Alexander Pritzel},
  \bibinfo{person}{Nicolas Heess}, \bibinfo{person}{Tom Erez},
  \bibinfo{person}{Yuval Tassa}, \bibinfo{person}{David Silver}, {and}
  \bibinfo{person}{Daan Wierstra}.} \bibinfo{year}{2015}\natexlab{}.
\newblock \showarticletitle{Continuous control with deep reinforcement
  learning}.
\newblock \bibinfo{journal}{\emph{arXiv preprint arXiv:1509.02971}}
  (\bibinfo{year}{2015}).
\newblock


\bibitem[\protect\citeauthoryear{Lowe, Wu, Tamar, Harb, Pieter~Abbeel, and
  Mordatch}{Lowe et~al\mbox{.}}{2017}]%
        {MADDPG}
\bibfield{author}{\bibinfo{person}{Ryan Lowe}, \bibinfo{person}{Yi~I Wu},
  \bibinfo{person}{Aviv Tamar}, \bibinfo{person}{Jean Harb},
  \bibinfo{person}{OpenAI Pieter~Abbeel}, {and} \bibinfo{person}{Igor
  Mordatch}.} \bibinfo{year}{2017}\natexlab{}.
\newblock \showarticletitle{Multi-agent actor-critic for mixed
  cooperative-competitive environments}.
\newblock \bibinfo{journal}{\emph{Advances in neural information processing
  systems}}  \bibinfo{volume}{30} (\bibinfo{year}{2017}).
\newblock


\bibitem[\protect\citeauthoryear{Mao, Gong, Zhang, Xiao, and Ni}{Mao
  et~al\mbox{.}}{2019}]%
        {mao2019learning}
\bibfield{author}{\bibinfo{person}{Hangyu Mao}, \bibinfo{person}{Zhibo Gong},
  \bibinfo{person}{Zhengchao Zhang}, \bibinfo{person}{Zhen Xiao}, {and}
  \bibinfo{person}{Yan Ni}.} \bibinfo{year}{2019}\natexlab{}.
\newblock \showarticletitle{Learning multi-agent communication under
  limited-bandwidth restriction for internet packet routing}.
\newblock \bibinfo{journal}{\emph{arXiv preprint arXiv:1903.05561}}
  (\bibinfo{year}{2019}).
\newblock


\bibitem[\protect\citeauthoryear{Niu, Paleja, and Gombolay}{Niu
  et~al\mbox{.}}{2021}]%
        {magic}
\bibfield{author}{\bibinfo{person}{Yaru Niu}, \bibinfo{person}{Rohan~R Paleja},
  {and} \bibinfo{person}{Matthew~C Gombolay}.} \bibinfo{year}{2021}\natexlab{}.
\newblock \showarticletitle{Multi-Agent Graph-Attention Communication and
  Teaming.}. In \bibinfo{booktitle}{\emph{AAMAS}}. \bibinfo{pages}{964--973}.
\newblock


\bibitem[\protect\citeauthoryear{Oliehoek and Amato}{Oliehoek and
  Amato}{2016}]%
        {pomdp}
\bibfield{author}{\bibinfo{person}{Frans~A Oliehoek} {and}
  \bibinfo{person}{Christopher Amato}.} \bibinfo{year}{2016}\natexlab{}.
\newblock \bibinfo{booktitle}{\emph{A concise introduction to decentralized
  {POMDPs}}}.
\newblock \bibinfo{publisher}{Springer}.
\newblock


\bibitem[\protect\citeauthoryear{Peng, Wen, Yang, Yuan, Tang, Long, and
  Wang}{Peng et~al\mbox{.}}{2017}]%
        {BiCNet}
\bibfield{author}{\bibinfo{person}{Peng Peng}, \bibinfo{person}{Ying Wen},
  \bibinfo{person}{Yaodong Yang}, \bibinfo{person}{Quan Yuan},
  \bibinfo{person}{Zhenkun Tang}, \bibinfo{person}{Haitao Long}, {and}
  \bibinfo{person}{Jun Wang}.} \bibinfo{year}{2017}\natexlab{}.
\newblock \showarticletitle{Multiagent bidirectionally-coordinated nets:
  Emergence of human-level coordination in learning to play starcraft combat
  games}.
\newblock \bibinfo{journal}{\emph{arXiv preprint arXiv:1703.10069}}
  (\bibinfo{year}{2017}).
\newblock


\bibitem[\protect\citeauthoryear{Rangwala and Williams}{Rangwala and
  Williams}{2020}]%
        {sarnet}
\bibfield{author}{\bibinfo{person}{Murtaza Rangwala} {and}
  \bibinfo{person}{Ryan Williams}.} \bibinfo{year}{2020}\natexlab{}.
\newblock \showarticletitle{Learning multi-agent communication through
  structured attentive reasoning}.
\newblock \bibinfo{journal}{\emph{Advances in Neural Information Processing
  Systems}}  \bibinfo{volume}{33} (\bibinfo{year}{2020}),
  \bibinfo{pages}{10088--10098}.
\newblock


\bibitem[\protect\citeauthoryear{Rashid, Samvelyan, Schroeder, Farquhar,
  Foerster, and Whiteson}{Rashid et~al\mbox{.}}{2018}]%
        {QMIX}
\bibfield{author}{\bibinfo{person}{Tabish Rashid}, \bibinfo{person}{Mikayel
  Samvelyan}, \bibinfo{person}{Christian Schroeder}, \bibinfo{person}{Gregory
  Farquhar}, \bibinfo{person}{Jakob Foerster}, {and} \bibinfo{person}{Shimon
  Whiteson}.} \bibinfo{year}{2018}\natexlab{}.
\newblock \showarticletitle{{QMIX}: Monotonic value function factorisation for
  deep multi-agent reinforcement learning}. In
  \bibinfo{booktitle}{\emph{International conference on machine learning}}.
  PMLR, \bibinfo{pages}{4295--4304}.
\newblock


\bibitem[\protect\citeauthoryear{Ruan, Du, Xiong, Xing, Li, Meng, Zhang, Wang,
  and Xu}{Ruan et~al\mbox{.}}{2022}]%
        {gcs}
\bibfield{author}{\bibinfo{person}{Jingqing Ruan}, \bibinfo{person}{Yali Du},
  \bibinfo{person}{Xuantang Xiong}, \bibinfo{person}{Dengpeng Xing},
  \bibinfo{person}{Xiyun Li}, \bibinfo{person}{Linghui Meng},
  \bibinfo{person}{Haifeng Zhang}, \bibinfo{person}{Jun Wang}, {and}
  \bibinfo{person}{Bo Xu}.} \bibinfo{year}{2022}\natexlab{}.
\newblock \showarticletitle{GCS: Graph-Based Coordination Strategy for
  Multi-Agent Reinforcement Learning}.
\newblock \bibinfo{journal}{\emph{arXiv preprint arXiv:2201.06257}}
  (\bibinfo{year}{2022}).
\newblock


\bibitem[\protect\citeauthoryear{Samvelyan, Rashid, De~Witt, Farquhar,
  Nardelli, Rudner, Hung, Torr, Foerster, and Whiteson}{Samvelyan
  et~al\mbox{.}}{2019}]%
        {smac}
\bibfield{author}{\bibinfo{person}{Mikayel Samvelyan}, \bibinfo{person}{Tabish
  Rashid}, \bibinfo{person}{Christian~Schroeder De~Witt},
  \bibinfo{person}{Gregory Farquhar}, \bibinfo{person}{Nantas Nardelli},
  \bibinfo{person}{Tim~GJ Rudner}, \bibinfo{person}{Chia-Man Hung},
  \bibinfo{person}{Philip~HS Torr}, \bibinfo{person}{Jakob Foerster}, {and}
  \bibinfo{person}{Shimon Whiteson}.} \bibinfo{year}{2019}\natexlab{}.
\newblock \showarticletitle{The starcraft multi-agent challenge}.
\newblock \bibinfo{journal}{\emph{arXiv preprint arXiv:1902.04043}}
  (\bibinfo{year}{2019}).
\newblock


\bibitem[\protect\citeauthoryear{Sartoretti, Kerr, Shi, Wagner, Kumar, Koenig,
  and Choset}{Sartoretti et~al\mbox{.}}{2019}]%
        {sartoretti2019primal}
\bibfield{author}{\bibinfo{person}{Guillaume Sartoretti},
  \bibinfo{person}{Justin Kerr}, \bibinfo{person}{Yunfei Shi},
  \bibinfo{person}{Glenn Wagner}, \bibinfo{person}{TK~Satish Kumar},
  \bibinfo{person}{Sven Koenig}, {and} \bibinfo{person}{Howie Choset}.}
  \bibinfo{year}{2019}\natexlab{}.
\newblock \showarticletitle{Primal: Pathfinding via reinforcement and imitation
  multi-agent learning}.
\newblock \bibinfo{journal}{\emph{IEEE Robotics and Automation Letters}}
  \bibinfo{volume}{4}, \bibinfo{number}{3} (\bibinfo{year}{2019}),
  \bibinfo{pages}{2378--2385}.
\newblock


\bibitem[\protect\citeauthoryear{Shalev-Shwartz, Shammah, and
  Shashua}{Shalev-Shwartz et~al\mbox{.}}{2016}]%
        {shalev2016safe}
\bibfield{author}{\bibinfo{person}{Shai Shalev-Shwartz},
  \bibinfo{person}{Shaked Shammah}, {and} \bibinfo{person}{Amnon Shashua}.}
  \bibinfo{year}{2016}\natexlab{}.
\newblock \showarticletitle{Safe, multi-agent, reinforcement learning for
  autonomous driving}.
\newblock \bibinfo{journal}{\emph{arXiv preprint arXiv:1610.03295}}
  (\bibinfo{year}{2016}).
\newblock


\bibitem[\protect\citeauthoryear{Singh, Jain, and Sukhbaatar}{Singh
  et~al\mbox{.}}{2019}]%
        {ic3}
\bibfield{author}{\bibinfo{person}{Amanpreet Singh}, \bibinfo{person}{Tushar
  Jain}, {and} \bibinfo{person}{Sainbayar Sukhbaatar}.}
  \bibinfo{year}{2019}\natexlab{}.
\newblock \showarticletitle{Individualized controlled continuous communication
  model for multiagent cooperative and competitive tasks}. In
  \bibinfo{booktitle}{\emph{International conference on learning
  representations}}.
\newblock


\bibitem[\protect\citeauthoryear{Sukhbaatar, Fergus, et~al\mbox{.}}{Sukhbaatar
  et~al\mbox{.}}{2016}]%
        {CommNet}
\bibfield{author}{\bibinfo{person}{Sainbayar Sukhbaatar}, \bibinfo{person}{Rob
  Fergus}, {et~al\mbox{.}}} \bibinfo{year}{2016}\natexlab{}.
\newblock \showarticletitle{Learning multiagent communication with
  backpropagation}.
\newblock \bibinfo{journal}{\emph{Advances in neural information processing
  systems}}  \bibinfo{volume}{29} (\bibinfo{year}{2016}).
\newblock


\bibitem[\protect\citeauthoryear{Sutton, McAllester, Singh, and Mansour}{Sutton
  et~al\mbox{.}}{1999}]%
        {sutton1999policy}
\bibfield{author}{\bibinfo{person}{Richard~S Sutton}, \bibinfo{person}{David
  McAllester}, \bibinfo{person}{Satinder Singh}, {and} \bibinfo{person}{Yishay
  Mansour}.} \bibinfo{year}{1999}\natexlab{}.
\newblock \showarticletitle{Policy gradient methods for reinforcement learning
  with function approximation}.
\newblock \bibinfo{journal}{\emph{Advances in neural information processing
  systems}}  \bibinfo{volume}{12} (\bibinfo{year}{1999}).
\newblock


\bibitem[\protect\citeauthoryear{Tan}{Tan}{1993}]%
        {IQL}
\bibfield{author}{\bibinfo{person}{Ming Tan}.} \bibinfo{year}{1993}\natexlab{}.
\newblock \showarticletitle{Multi-agent reinforcement learning: Independent vs.
  cooperative agents}. In \bibinfo{booktitle}{\emph{Proceedings of the tenth
  international conference on machine learning}}. \bibinfo{pages}{330--337}.
\newblock


\bibitem[\protect\citeauthoryear{Vaswani, Shazeer, Parmar, Uszkoreit, Jones,
  Gomez, Kaiser, and Polosukhin}{Vaswani et~al\mbox{.}}{2017}]%
        {Attention}
\bibfield{author}{\bibinfo{person}{Ashish Vaswani}, \bibinfo{person}{Noam
  Shazeer}, \bibinfo{person}{Niki Parmar}, \bibinfo{person}{Jakob Uszkoreit},
  \bibinfo{person}{Llion Jones}, \bibinfo{person}{Aidan~N Gomez},
  \bibinfo{person}{{\L}ukasz Kaiser}, {and} \bibinfo{person}{Illia
  Polosukhin}.} \bibinfo{year}{2017}\natexlab{}.
\newblock \showarticletitle{Attention is all you need}.
\newblock \bibinfo{journal}{\emph{Advances in neural information processing
  systems}}  \bibinfo{volume}{30} (\bibinfo{year}{2017}).
\newblock


\bibitem[\protect\citeauthoryear{Veli{\v{c}}kovi{\'c}, Cucurull, Casanova,
  Romero, Lio, and Bengio}{Veli{\v{c}}kovi{\'c} et~al\mbox{.}}{2017}]%
        {GAT}
\bibfield{author}{\bibinfo{person}{Petar Veli{\v{c}}kovi{\'c}},
  \bibinfo{person}{Guillem Cucurull}, \bibinfo{person}{Arantxa Casanova},
  \bibinfo{person}{Adriana Romero}, \bibinfo{person}{Pietro Lio}, {and}
  \bibinfo{person}{Yoshua Bengio}.} \bibinfo{year}{2017}\natexlab{}.
\newblock \showarticletitle{Graph attention networks}.
\newblock \bibinfo{journal}{\emph{arXiv preprint arXiv:1710.10903}}
  (\bibinfo{year}{2017}).
\newblock


\bibitem[\protect\citeauthoryear{Zhang, Zhang, and Lin}{Zhang
  et~al\mbox{.}}{2019}]%
        {VBC}
\bibfield{author}{\bibinfo{person}{Sai~Qian Zhang}, \bibinfo{person}{Qi Zhang},
  {and} \bibinfo{person}{Jieyu Lin}.} \bibinfo{year}{2019}\natexlab{}.
\newblock \showarticletitle{Efficient communication in multi-agent
  reinforcement learning via variance based control}.
\newblock \bibinfo{journal}{\emph{Advances in Neural Information Processing
  Systems}}  \bibinfo{volume}{32} (\bibinfo{year}{2019}).
\newblock


\end{thebibliography}

\clearpage

\appendix
\section{Training Details}
This section introduces the hyperparameters of the policy optimization algorithms and network architecture. Hyperparameters used in the DDPG and REINFORCE algorithms are given in Table \ref{table: hyperparameter}. 

\begin{table}[h]
	\centering
	\caption{Hyperarameters settings}

	\begin{tabular}{cc}
        Parameter & Value\\
        \hline
        Policy (Actor) Learning Rate & $1 \times 10^{-4}$\\
        Critic Learning Rate & $1 \times 10^{-4}$ \\
        Gradient Norm Clipping & 0.1\\
        Optimizer & Adam \citep{adam} \\
        Discount Rate $\gamma$ & 0.99 \\
        Target Network Update Period & 200\\
        Soft Update Value $\tau$ & 0.01\\
        Episode Length & 50, 60 or 80\\
        \hline
    \end{tabular}
    \label{table: hyperparameter}
\end{table}

The specifics of the actor's architecture are shown in Table \ref{table: network_details}. Note that the critic network has similar architecture as the actor network except for the output layer.

\begin{table}[h]
	\centering
	\caption{Architecture settings}

	\begin{tabular}{cc}
        Network & Actor\\
        \hline
        Encoder size & 128\\
        Hidden Embedding size & 128 \\
        Query Size & 128\\
        key Size & 128\\
        Value Size & 128\\
        First-round communication embedding size & 128\\
        Second-round Communication embedding size & 128\\ 
        Policy network embedding & [384, 128]\\
        \hline
    \end{tabular}
    \label{table: network_details}
\end{table}

The value of threshold T we set for the controller is domain-dependent. The specific values are listed in the Table \ref{table: threshold_value}.

\begin{table}[h]
	\centering
	\caption{The value of threshold $T$}

	\begin{tabular}{cc}
        Scenario & Value\\
        \hline
        Cooperative Navigation & from 0.1 to 0.6\\
        Predator Prey & from 0.1 to 0.6\\
        Traffic Junction & from 0.1 to 0.2\\
        \hline
    \end{tabular}
    \label{table: threshold_value}
\end{table}

All experiments are conducted with 5 different seeds on Nvidia RTX A5000. 

\section{Environment Settings}

In order to make our proposed method reproducible, we introduce the details of the environment settings. 

\subsection{Cooperative Navigation}
Our implementation of Cooperative Navigation is based on the official multi-particle environment release \citep{MADDPG}. In order to highlight the characteristics of partial observation, we follow \citep{i2c} and modify the original observation and reward function: the revised observation contains current agent's position, current agent's velocity, all landmarks' positions and the two closest agents' positions. The reward function is revised as $\text{original reward}\times0.01 + \text{detection reward}$. The detection reward is given as follows:
If the distance between the agent and its nearest landmark is less than 0.2, it will get a reward of +0.5, and if the distance is less than 0.1, it will get a reward of +1. Besides, if the distance between 2 agents is less than 0.5, each of them will receive a penalty -0.25. 

\subsection{Predator Prey}
Following the previous work \citep{i2c,ic3}, we adopt the cooperative version of predator prey environment. Treating the preys as landmarks, we set the reward function in the same way as cooperative navigation. And the partial observation contains current agent's position, current agent's velocity, three closest landmarks' positions and three closest agents' positions.

\subsection{Traffic Junction}
\begin{figure}[h]
\centering
\subfigure[Medium Mode]{
\includegraphics[width =  0.2\textwidth]{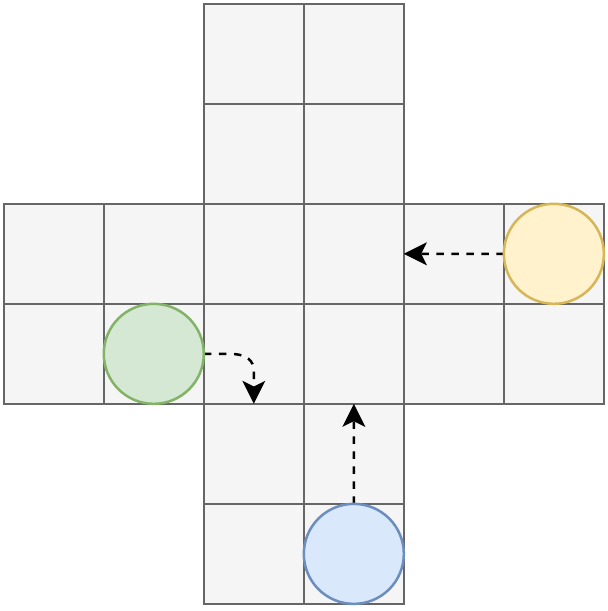}
\label{fig:medium}
}
\subfigure[Hard Mode]{
\includegraphics[width =  0.2\textwidth]{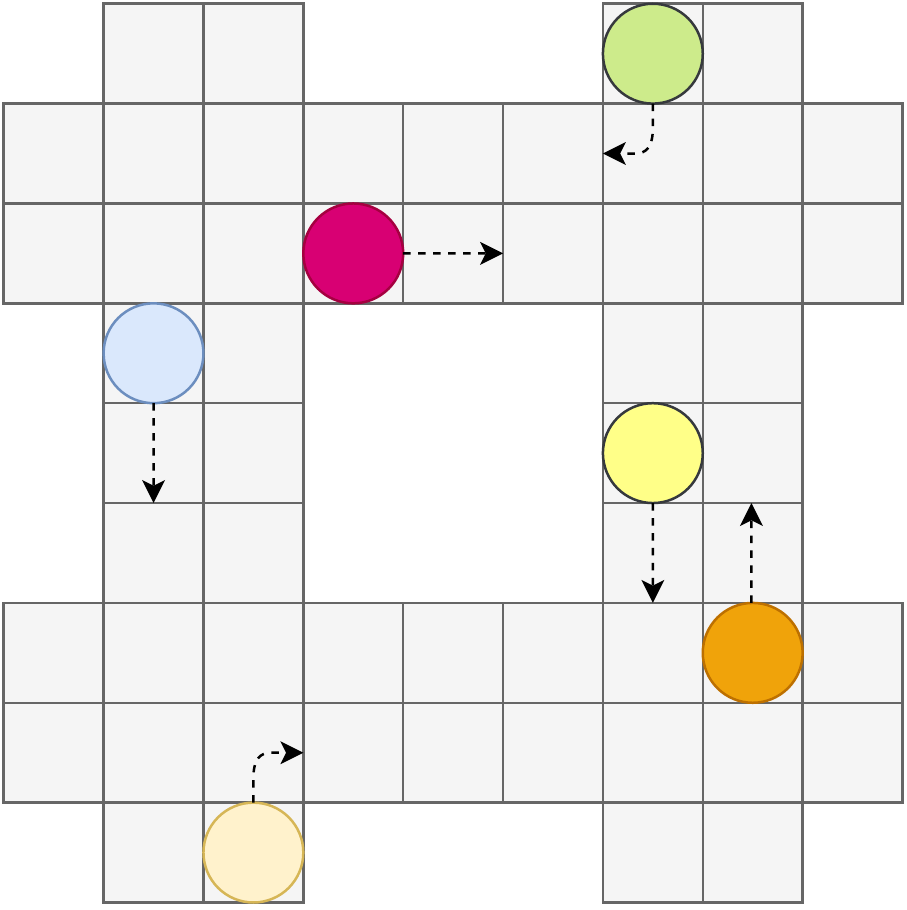}
\label{fig:hard}
}

\caption{Traffic Junction
}
\label{Fig_tj}
\end{figure}

In this environment, we adopt the same observation and reward function as \citep{i2c, sarnet}. In order to emphasize the effect of communication, we set each agent's vision range to 0, which means they can only access other cars' information by communication. The detailed environment settings are provided in Table \ref{table: traffic_junction settings} and Figure \ref{Fig_tj}. Besides, we compute the successful rate in the same way as \citep{i2c}: if no collision happens in an episode, then we regard it as a success episode.

\begin{table}[h]
	\centering
	\caption{Traffic junction environment settings}

	\begin{tabular}{ccc}
        Difficulty & Medium & Hard\\
        \hline
        P-arrive & 0.2 & 0.2\\
        Max Number of Agents & 10 & 20 \\
        Entries & 8 & 16\\
        Routes & 4 & 8\\
        Junction(s) & 1 & 4\\
        World Dimension & $6\times6$ & $9\times9$\\
        \hline
    \end{tabular}
    \label{table: traffic_junction settings}
\end{table}

\section{Additional Experiments}

\begin{figure}[H] 
\centering 
\includegraphics[width=5cm]{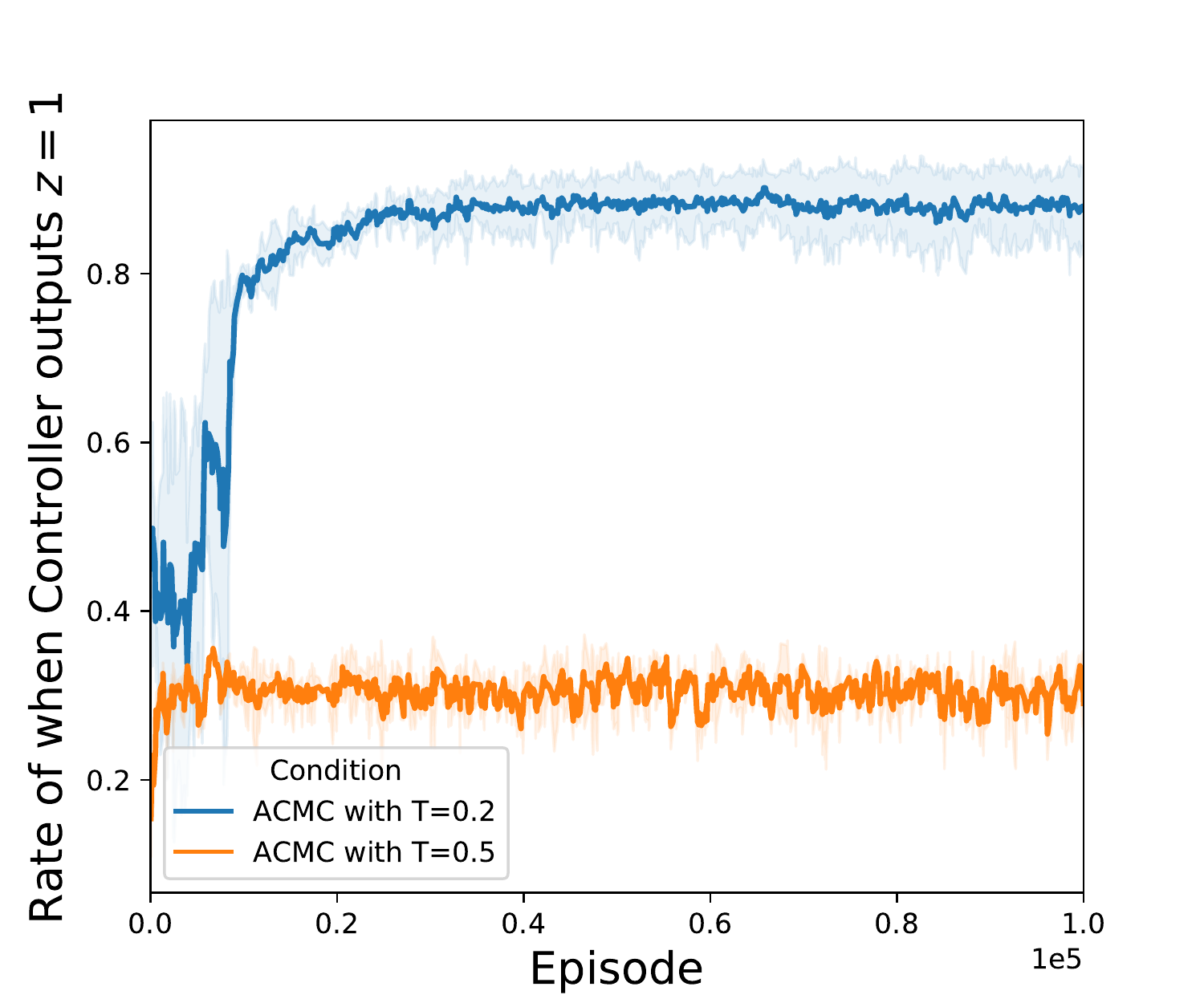} 
\caption{The opening rate of Controller under different threshold $T$.
} 
\label{cutted_links.pdf} 
\end{figure}

In this section, we provide more details about the controller. We compare the opening rate of the controller when $T=0.2$ and $T=0.5$. The specific values are given in Figure \ref{cutted_links.pdf}. 

\end{document}